\documentclass[11pt]{article}
\pdfoutput=1
\usepackage{jcappub,natbib,float,caption,comment,bm}

\def\be{\begin{equation}}
\def\ee{\end{equation}}
\def\ba#1\ea{\begin{align}#1\end{align}}
\newcommand{\vs}{\nonumber\\}

\renewcommand{\v}[1]{\mathbf{#1}}
\newcommand{\vx}{\v{x}}

\newcommand{\refeq}[1]{Eq.~(\ref{eq:#1})}          
\newcommand{\refeqs}[2]{Eqs.~(\ref{eq:#1})--(\ref{eq:#2})}          
\newcommand{\reffig}[1]{figure~\ref{fig:#1}}          
\newcommand{\refFig}[1]{Figure~\ref{fig:#1}}
\newcommand{\refFigs}[2]{Figures~(\ref{fig:#1})--(\ref{fig:#2})}          
\newcommand{\refsec}[1]{section~\ref{sec:#1}}          
\newcommand{\refSec}[1]{Section~\ref{sec:#1}}          
\newcommand{\refapp}[1]{Appendix~\ref{app:#1}}

\newcommand{\Om}{\Omega_m}

\newcommand{\OL}{\Omega_\Lambda}

\renewcommand{\d}{\delta}

\newcommand{\D}{\Delta}
\newcommand{\dr}{\delta_\rho}
\newcommand{\s}[1]{\sigma_{#1}}
\newcommand{\T}{{\rm TH}}
\newcommand{\G}{{\rm G}}

\def\Mpch{\,h^{-1}{\rm Mpc}}

\newcommand{\<}{\left\langle}
\renewcommand{\>}{\right\rangle}

\title{Precision measurement of the local bias of dark matter halos}

\author[a]{Titouan Lazeyras,}
\author[a,b]{Christian~Wagner,}
\author[c]{Tobias Baldauf,}
\author[a]{Fabian~Schmidt}
\affiliation[a]{Max-Planck-Institut f\"ur Astrophysik, Karl-Schwarzschild-Str. 1, 85748 Garching, Germany}
\affiliation[b]{Leibniz-Institut f\"ur Astrophysik, An der Sternwarte 16, 14482 Potsdam, Germany}
\affiliation[c]{Institute for Advanced Study, Einstein Drive, Princeton, NJ 08540, United States}
\emailAdd{titouan@mpa-garching.mpg.de, cwagner@mpa-garching.mpg.de,
t.baldauf@tbaweb.de, fabians@mpa-garching.mpg.de}
\abstract{
We present accurate measurements of the linear, quadratic, and cubic local bias of dark matter halos, using curved ``separate universe'' N-body simulations which effectively incorporate an infinite-wavelength overdensity.  This can be seen as an exact implementation of the peak-background split argument.  We compare the
results with the linear and quadratic bias measured from the halo-matter
power spectrum and bispectrum, and find good agreement.  On the other hand,
the standard peak-background split applied to the Sheth \& Tormen (1999) and 
Tinker et al. (2008) halo mass functions matches the measured linear bias parameter only at the level of 10\%.  The prediction from the excursion set-peaks approach performs much better, which  
can be attributed to the stochastic moving barrier employed in the excursion set-peaks
prediction.  We also provide convenient fitting formulas for the nonlinear bias
parameters $b_2(b_1)$ and $b_3(b_1)$, which work well over a range
of redshifts.}

\begin{document}
\maketitle
\flushbottom

%%%%%%%%%%%%%%%%%%%%%%%%%%%%%%%%%%%%%%%%%%%%%%%%%%%%%%%%%%%%%%%%%%%%%%%%%%%%%
%%%%%%%%%%%%%%%%%%%%%%%%%%%%%%%%%%%%%%%%%%%%%%%%%%%%%%%%%%%%%%%%%%%%%%%%%%%%%
\section{Introduction}
\label{sec:intro}

The large-scale distribution of dark matter halos is one of the key
ingredients of the theoretical description of large-scale structure (LSS).  
Since most observed tracers of LSS, such as galaxies, reside in halos,
the statistics of halos determine those of galaxies on large scales.  
Similarly, the halo model description of the nonlinear matter density
field \cite{cooray/sheth} crucially relies on halo statistics.  In the context of perturbation
theory, the statistics of halos are written in terms of bias parameters 
multiplying operators constructed out of the matter density field.  
In general, these operators consist of powers of the matter density
and tidal field \cite{fry/gaztanaga,mcdonald/roy}, as well as convective time derivatives of these quantities \cite{MSZ,senatore:14}.  
However, the most well-studied and phenomenologically most important
bias parameters on large scales are those multiplying powers of
the matter density field, i.e. 
\be
\d_h(\vx,\tau) \supset b_1(\tau) \dr(\vx,\tau) + \frac12 b_2(\tau) \dr^2(\vx,\tau)
+ \frac16 b_3(\tau) \dr^3(\vx,\tau) + \cdots\,,
\label{eq:localbias}
\ee
where $\d_h$ is the fractional number density perturbation of a given
halo sample,
while $\dr$ is the matter density perturbation.  More precisely,
the powers of $\dr$ should be understood as renormalized operators
\cite{mcdonald,assassi/etal,PBSpaper}.  
The $b_n$ are commonly called (nonlinear) \emph{local bias parameters}.  
The goal of this paper is to present precision measurements of
$b_1,\,b_2,\,b_3$ using a novel technique, \emph{separate universe simulations}.

In the separate universe approach 
\citep{lemaitre:1933,sirko:2005,baldauf/etal:2011,li/hu/takada:2014,Wagner:2014}, a long-wavelength density
perturbation is included in an N-body simulation by changing the 
cosmological parameters, in particular $\Om,\,\OL,\,\Omega_K$ and $H_0$, 
from their fiducial values, and running the simulation to a different
scale factor.  As argued in \cite{baldauf/etal:11,jeong/etal,PBSpaper}, 
the (renormalized) local bias parameters defined in \refeq{localbias}
correspond to the response of the halo abundance, $\bar n_h$, to a long-wavelength
density perturbation, equivalent to a change in the background density, $\bar\rho$,
\be
b_{n} = \frac{\bar\rho^{\hskip 1pt n}}{\bar n_h} \frac{\partial^n \bar n_h}{\partial\bar\rho^{\hskip 1pt n}}\, .
\ee
This can be understood as an exact formulation of the peak-background split (PBS) \cite{kaiser:1984,mo/white:1996}.  Thus, the $b_n$ can be measured through the mass function of halos in a suite
of separate universe simulations.  This technique has several advantages:  
first, it is guaranteed to recover the large-scale limit of the $b_n$, without
scale-dependent or nonlinear corrections which affect measurements of the
bias parameters from the halo power spectrum and bispectrum, or
from the cross-correlation with smoothed fields.  Note that, starting 
at second order, ``nonlocal'' bias parameters such as those with respect
to powers of the tidal field will enter in these latter measurements at
the same level as the $b_n$.  Second, 
we can straightforwardly obtain measurements of higher order bias parameters
such as $b_3$, which become cumbersome to measure using correlations.  Finally,
by using the same initial phases for simulations with different density,
we can cancel to a large extent the cosmic variance contribution to the measurement error.  

Separate universe simulations are expected to estimate the same set of bias parameters as those obtained from matter-halo cross-correlations. We will thus compare the biases obtained from the separate universe simulations to
those determined by fitting to halo two- and three-point statistics.  
We also compare the results to biases derived from universal mass functions
using the classic peak-background split argument, and 
recent theoretical predictions from
the excursion set-peaks (ESP) approach \cite{Paranjape:2012,Paranjape:2013}, which incorporates some aspects
of the Gaussian peaks model into the excursion set framework.  

Higher order bias parameters have previously been measured in simulations
by correlating the halo number with powers of the smoothed density field 
at the final time (Eulerian frame) 
\cite{angulo/baugh/lacey:2008,manera/gaztanaga:2011}
or in the initial conditions \cite{paranjape/etal:2013}.  
However, the bias parameters measured in this way depend on the smoothing
scale adopted, while the local bias parameters that are relevant for perturbation theory predictions, and that we are interested in here, correspond to
a smoothing scale of infinity.  Further, all these references
neglect the nonlocal bias terms mentioned above, 
which will affect the inferred values of $b_2$ and higher.  
For these reasons, it is difficult to directly compare our measurements of
nonlinear bias parameters with these previous results (although we find
broad agreement).    
We stress again that in the separate universe approach we are guaranteed
to obtain the local bias in the large-scale limit, without nonlinear
or tidal corrections.  Moreover, we simultaneously obtain both
the Eulerian ($b_n$) and Lagrangian ($b_n^L$) bias parameters.  

Two related papers appeared on the preprint archive simultaneously
to this paper.  Ref.~\cite{li/etal:15} measured the linear bias using
separate universe simulations through an abundance matching technique
which yields the integrated halo bias above a mass threshold.  This
technique reduces the shot noise in the bias measurement.  
Ref.~\cite{baldauf/etal:15} also measured the linear bias via
the mass function.  In addition, they present measurements of $b_2$
through the response of the halo power spectrum to a long-wavelength mode
(as done in \cite{li/hu/takada:2014,Wagner:2015} for the matter power spectrum).  
Our results are consistent with the findings of both of these references.  
However, unlike these and any other previous published results, we
use the fully nonlinear separate universe approach to obtain accurate
measurements of the \emph{linear and nonlinear} local biases.

In this paper we adopt a flat $\Lambda$CDM fiducial cosmology with $\Om=0.27$, $h=0.7$, $\Omega_b h^2=0.023$ and $\mathcal{A}_s=2.2\cdot 10^{-9}$.  
The outline of the paper is as follows.  In \refsec{theory}, we present
the theoretical predictions that we will compare our measurements with.  \refSec{bsep}
describes the technique of measuring bias parameters from separate universe
simulations, while \refsec{bcorr} presents the estimators for $b_1$ and $b_2$
using the conventional approach of measuring halo correlations.  
We discuss the results in \refsec{res}.  We conclude in \refsec{concl}.  
The appendices contain more details on the ESP predictions as well as
our bias measurements.  

%%%%%%%%%%%%%%%%%%%%%%%%%%%%%%%%%%%%%%%%%%%%%%%%%%%%%%%%%%%%%%%%%%%%%%%%%%%%%
%%%%%%%%%%%%%%%%%%%%%%%%%%%%%%%%%%%%%%%%%%%%%%%%%%%%%%%%%%%%%%%%%%%%%%%%%%%%%
\section{Theory predictions}
\label{sec:theory}

In this section we present several theoretical predictions for the large-scale bias from the literature.  We first recap the PBS argument in \refsec{PBS} and briefly present the ESP formalism in \refsec{ESP}.

Before jumping into details, we briefly explain the definitions of \textit{Lagrangian} and \textit{Eulerian} halo bias. The Lagrangian bias links the abundance of dark matter halos to the density perturbations in Lagrangian space, i.e. it describes the relation of proto-halos in the initial conditions that correspond to halos identified at redshift $z$ to the initial linear density perturbation field. On the other hand, the Eulerian bias relates the halos identified at redshift $z$ to the nonlinear density field, $\dr$, at redshift $z$.  
In the case of the local bias parameters considered here, there is
an exact nonlinear mapping between the Lagrangian bias parameters $b_n^L$
and their Eulerian counterparts $b_m$, see \refapp{compbsep}.  We will
make use of this mapping both for the theory predictions and measurements.  

In the following, the top-hat filtered variance on a scale $R_\T$ 
(the Lagrangian radius of halos) is denoted as
\be
\s{0}^2 \equiv \int {\rm dln} k\, \D^2 (k) [W_\T(kR_\T)]^2,
\label{eq:sigma0}
\ee
where $\D^2(k) = k^3 P(k)/2\pi^2$ is the dimensionless linearly extrapolated matter power spectrum and 
the top-hat filter in Fourier space $W_\T(k R_\T)$ is given in \refeq{WTH}.  

\subsection{Peak-background split bias}
\label{sec:PBS}

We briefly recap how the bias parameters can be derived from the differential halo mass function using the PBS argument,
as initially proposed in \cite{kaiser:1984,cole/kaiser:1989,mo/white:1996}.  
Following the PBS argument, the effect of a long wavelength mode $\d_0$ on the small scale formation can be seen as locally modulating the density threshold for halo formation, or barrier $B$, sending it to $B-\d_0$ (here we denote the barrier as $B$ to emphasize that this argument is not restricted to the constant spherical collapse threshold $\d_c$ and can be extended to barriers depending e.g. on the halo mass $M$ through $\s{0}$). Note that, in the case where stochasticity should be introduced in the barrier, this shift does not modify the stochastic contribution to the barrier, which is supposed to capture the effect of small-scale modes. We define the differential mass function as  
\be
n_{\rm}(\nu_{\rm B}) = \frac{\bar{\rho}_m}{M}f_{\rm}(\nu_{\rm B}) \left|\frac{{\rm d ln}\,\s{0}}{{\rm d ln }\,M}\right|,
\label{eq:nf}
\ee
with $\nu_{\rm B} \equiv B(\s{0})/\s{0}$ (we reserve the notation $\nu$ for $\nu\equiv\d_c/\s{0}$), $M$ the corresponding mass and $f(\nu_{\rm B})$ the mass fraction contained in halos of mass $M$. The scale-independent large-scale Lagrangian bias parameters are then defined by the well known relation 
\be
b^L_n(\nu_{\rm B}) = \frac{1}{n(\nu_{\rm B})}\frac{\partial ^n n([B(\s{0})-\d_0]/\sigma_0)}{\partial \d_0^n}\Bigg|_{\d_0=0}.
\label{eq:biasPBS}
\ee
As we have indicated, this also applies if the deterministic part of the barrier is mass-dependent. We will use \refeq{biasPBS} both to derive the bias in the ESP model and from the fits to the mass function proposed in \cite{Sheth:1999} and \cite{Tinker:2008} (hereafter ST99 and T08 respectively). 

\subsection{Excursion set peaks}
\label{sec:ESP}

In this section, we review the ESP formalism proposed in \citep{Paranjape:2012} and \citep{Paranjape:2013}.  The details of the calculation are relegated to \refapp{ESP}.  All the results that we present here and in \refapp{ESP} were already derived in these two references, but in a different way; here, we use the PBS argument to derive the bias parameters directly.  Further, the ESP predictions
for $b_3$ and $b_4$ are computed here for the first time.  

The ESP aims at unifying the peak model of Bardeen et al. in 1986 (hereafter BBKS) \citep{Bardeen:1986} and the excursion set formalism of Bond et al. in 1991 \citep{Bond:1991}. It can be seen either as addressing the cloud-in-cloud problem within the peak model, or as applying the excursion set formalism to a special subset of all possible positions (the peaks).  
We follow \citep{Paranjape:2013}, who chose a top-hat filter for the excursion
set part, and a Gaussian filter to identify peaks (in order to ensure finite moments
of derivatives of the smoothed density field).  

More importantly, \citep{Paranjape:2013} improved the model by adding a mass-dependent stochastic scatter to the threshold.  Specifically, the barrier is
defined as \citep{Paranjape:2012}
\be 
B(\s{0}) = \d_c + \beta \s{0}\,.
\label{eq:barrier}
\ee
Here, $\beta$ is a stochastic variable and \cite{Paranjape:2013} chose its PDF $p(\beta)$ to be lognormal with mean and variance corresponding to $\<\beta\> = 0.5$ and ${\rm Var}(\beta)=0.25$.  This choice was made to match the peak height measured in simulations by \cite{robertson/etal}.   Hence $\beta$ takes only positive values.  Note that \refeq{barrier} then corresponds to a mass-dependent mean barrier $\d_c + 0.5 \s{0}$.  

As we show in \refapp{ESP}, the Lagrangian bias parameters in the ESP
can be directly derived from \refeq{biasPBS} by inserting the multiplicity
function $f_{\rm ESP}(\nu)$ into \refeq{nf}, and sending 
$\nu = \d_c/\s{0}$ to $\nu_1 = \nu\left(1-\d_0/\d_c\right)$.\footnote{Here one needs to take care not to shift one instance of $\nu$ in the expression for $f_{\rm ESP}(\nu)$ that is actually unrelated to the barrier. See \refapp{ESP}.}  
Our results for the bias, \refeq{btheo}, are identical to the large-scale bias parameters derived using
a different approach in \citep{Paranjape:2012,Paranjape:2013}.  
We will see that the choice of barrier \refeq{barrier} leads to significant differences from the standard PBS biases derived using $B = \d_c$
from the T08 and ST99 mass functions.

%%%%%%%%%%%%%%%%%%%%%%%%%%%%%%%%%%%%%%%%%%%%%%%%%%%%%%%%%%%%%%%%%%%%%%%%%%%%%
%%%%%%%%%%%%%%%%%%%%%%%%%%%%%%%%%%%%%%%%%%%%%%%%%%%%%%%%%%%%%%%%%%%%%%%%%%%%%
\section{Bias parameters from separate universe simulations}
\label{sec:bsep} 

Our results are based on the suite of separate universe simulations described in \cite{Wagner:2014,Wagner:2015}, performed using the cosmological code GADGET-2 \citep{Springel:2005}. The idea of the separate universe simulations is that a uniform matter overdensity $\dr$ of a scale larger than the simulation box can be absorbed in the background density $\tilde{\rho}_m$ of a modified cosmology simulation (throughout the whole paper, quantities in modified cosmologies will be denoted with a tilde), where
\be 
\tilde{\rho}_m(t) = \rho_m(t)\left[1+\dr(t)\right], 
\label{eq:dr}
\ee
with $\rho_m$ the mean matter density in a simulation with no overdensity (which we call the fiducial cosmology).  Indeed, a uniform density can only be included in this way, since the Poisson equation for the potential enforces a vanishing mean density perturbation over the entire box. Thus one can see a simulation with a constant overdensity $\dr$ as a separate universe simulation with a properly modified cosmology. Qualitatively, a positive overdensity causes slower expansion and enhances the growth of  structure, i.e. more halos, whereas a negative one will have the opposite effect. The precise mapping of $\dr$ to modified cosmological parameters is described in \cite{Wagner:2014}.  Crucially, we work to fully nonlinear order in $\dr(t)$.  

We use two sets of simulations denoted by ``lowres'' and ``highres'' throughout the paper. Both have a comoving box size of $500\Mpch$ in the fiducial cosmology.  The ``lowres'' set uses $256^3$ particles in each simulation, while ``highres'' employs $512^3$ particles.  For both sets, we run the fiducial cosmology, i.e. $\dr=0$,  and simulations with values of $\dr$ corresponding to $\d_L$ = \{$\pm$0.5, $\pm$0.4, $\pm$0.3, $\pm$0.2, $\pm$0.1, $\pm$0.07, $\pm$0.05, $\pm$0.02, $\pm$0.01\}, where $\d_L$ is the present-day linearly extrapolated matter density contrast. 
In addition, we simulate separate universe cosmologies corresponding to $\d_L$ = 0.15, 0.25, and 0.35 for both resolutions. 
This makes the sampling in the final, nonlinear $\dr$ more symmetric around 0 which should help diminish the covariance between the bias parameters.\footnote{We have not performed a systematic study on the number of $\d_L$ values that are necessary to derive accurate measurements of the $b_n$ up to a given order.  
Given the significant degeneracies between $b_n$ and $b_{n+2}$ we have found
(\refapp{cov}), this is a nontrivial question.}  
The comoving box size in the modified cosmology simulations is adjusted to match that in the fiducial cosmology, $L=500\Mpch$. 
Hence, in the high redshift limit ($z\rightarrow \infty$ for which $\dr\rightarrow 0$) the physical size of the box is the same for all simulations whereas at the present time ($z=0$ in the fiducial cosmology) the physical size of the simulation box varies with $\dr$. However, this choice of the box size has the advantage that the physical mass resolution is the same within each set of simulations regardless of the simulated overdensity $\dr$ (i.e. $\tilde{m}_p = m_p$ where $m_p$ is the particle mass in the fiducial cosmology). 
Since the biases are determined by comparing halo abundances between different overdensities, this eliminates any possible systematic effects in the biases due to varying mass resolution. 
The mass resolution is $m_p = 5.6\cdot 10^{11}h^{-1} M_\odot$ in the ``lowres'' set of simulations and $m_p=7\cdot 10^{10}h^{-1} M_\odot$ in the ``highres'' one. Furthermore, for the ``lowres'' set of simulation, we ran 64 realizations of the entire set of $\d_L$ values. For the ``highres'' one we ran only 16 realizations of each $\d_L$ value as they are more costly in terms of computation time.  
Each simulation was initialized using 2LPT at $z_i$ = 49.  
For further details about the simulations, see \citep{Wagner:2015}.

\subsection{Halo catalogs}
\label{sec:HC}

The halos were identified using the Amiga Halo Finder (hereafter AHF) \cite{Gill:2004,Knollmann:2009}, which identifies halos with a spherical overdensity (SO) algorithm.  We identify halos at a fixed proper time corresponding to $z=0$ in the fiducial cosmology.  
In this paper, we only use the number of distinct halos and do not consider their sub-halos. 

The key point in identifying halos with the spherical overdensity criterion is the setting of the density threshold.  We choose here a value of $\Delta_{\rm SO}=200$ times the background matter density in the \emph{fiducial} cosmology.  
Thus, our measured bias parameters are valid for this specific halo definition.  
For the simulations with a different background density, the threshold must be rescaled in order to compare halos identified using the same physical density in each simulation. Specifically, we need to use
\begin{equation}
\Delta_{\rm SO} = \frac{200}{1+\dr}\,.
\label{eq:DSO}
\end{equation}
Another point is the treatment of the particle unbinding in a halo.  AHF has the ability to remove unbound particles, i.e particles which are not gravitationally bound to the halo they are located in.  However, in order to avoid having to implement the complicated matching of the unbinding criterion between the modified and fiducial cosmologies, we have turned unbinding off in all halo catalogs.  
Note that the effect of unbinding is very small (of order 1\% on the mass function), and
that we consistently use the same halo catalogs for all measurements,
so that this choice does not affect our comparison between different methods
for measuring bias.  

We count halos in top-hat bins given by 
\be
W_n(M,M_{{\rm center}})=\begin{cases} 1 &\mbox{if } \left|{\rm log}_{10}(M)-{\rm log}_{10}(M_{{\rm center}})\right| \leq 0.1 \\
0 & \mbox{otherwise, } \end{cases}
\label{eq:bins}
\ee
where $M$ is the mass ($M_{{\rm center}}$ corresponding to center of the bin).  
For the high resolution simulations, we count halos in 12 bins centered from ${\rm log}_{10}\left(M_{{\rm center}}\right) = 12.55$ to ${\rm log}_{10}\left(M_{{\rm center}}\right) = 14.75$, to ensure that we have enough halos in each bin. For the low resolution simulations, we have 7 bins from ${\rm log}_{10}\left(M_{{\rm center}}\right) = 13.55$ to ${\rm log}_{10}\left(M_{{\rm center}}\right) = 14.75$. With this binning choice, the lowest bin is centered around halos with 63 particles for the ``lowres'' set of simulations, with a lower limit at halos containing around 50 particles. For the ``highres'' set of simulations, the lowest mass bin is centered on halos with around 51 particles, with a lower limit around 40 particles. These numbers are quite low compared to more conservative values (e.g. 400 particles in T08). However $\d_h$ is the \emph{relative difference} of the number of halos between the fiducial and modified cosmology simulations (see \refeq{DN_N} hereafter) and therefore that quantity should be less affected by resolution effects. For halos with a minimum number of 40 particles, we did not find any systematic difference between the bias parameters measured from the ``lowres'' and ``highres'' simulations. Thus, we present results for halos that are resolved by at least 40 particles. 

\subsection{Eulerian biases}
\label{sec:BE}

Instead of fitting the Eulerian bias parameters directly to the simulation results, we derive them from the measured Lagrangian biases for which the fitting is more robust, using the exact nonlinear evolution of $\dr$ (see \refapp{compbsep} for the details of the mapping).  
In order to obtain the Lagrangian bias parameters, we compute $\d_h(M,\d_L)$ versus $\d_L$ where $\d_h(M,\d_L)$ is the overdensity of halos in a bin of mass $M$ compared to the fiducial case $\d_L=0$,
\be
\delta_h (M,\d_L) = \frac{\tilde{N}(M,\d_L) - N(M)}{N(M)},
\label{eq:DN_N}
\ee
with $\tilde N(M,\d_L)$ the number of halos in a bin centered around mass $M$ in the presence of the linear overdensity $\d_L$ and $N(M)=\tilde N(M,\d_L=0)$. 
Note that $\d_h(M,\d_L)$ is the overdensity of halos in Lagrangian space as the physical volumes of the separate universe simulations only coincide at high redshift.

In order to obtain the Lagrangian bias parameters $b_n^L$, we then fit \refeq{DN_N} by  
\be
\d_h = \sum_{n=1}^5  \frac{1}{n!} b_n^L (\d_L)^n\,.
\label{eq:BiasExp}
\ee
As indicated in \refeq{BiasExp} we use a $5^{\rm th}$ order polynomial in $\d_L$ by default.  In \refapp{deg} we study the effect of the degree of the polynomial on the results; as a rough rule, if one is interested in $b_n^L$, then one should fit a polynomial up to order $n+2$.  

In order to estimate the overall best-fit of and error bars on the bias parameters, we use a bootstrap technique. For each non zero $\d_L$ value, we randomly produce $p$ resamples of the mass function. Each resample is composed of the same number of realizations as the original sample (i.e. 16 or 64) and we choose $p=100\cdot 64$ ($100\cdot 16$) for the low (high) resolution simulations. We then compute the average number of halos per mass bin for each resample. This gives us $p$ numbers $\tilde{N}^i(M,\d_L)$. 
For a given $\d_L$, we also create the same set of resamples for the fiducial cosmology  
and again compute the average number of halos, i.e. $N^i(M)$.  We then compute $p$ times $\d^i_h$ according to \refeq{DN_N} for every $\d_L$ value. 
Since we use the same resamples for the separate universe results, $\tilde{N}^i(M,\d_L)$, and the fiducial case, $N^i(M)$, the cosmic variance is canceled to leading order. The error on $\d_h$ at fixed mass and $\d_L$ is given by the sample variance and we use it as a weight for the fit. 
We neglect, however, the covariance between $\tilde{N}^i(M,\d_L)$ for different $\d_L$ values.
We then produce $p$ fits with a weighted least squares method. For every bias parameter, the value we report is the mean of the results of the $p$ fits while the corresponding error bar is given by the square root of the variance of the distribution.   Within the mass range common to both sets of simulations ``lowres'' and ``highres'', the measurements are consistent with each other and hence we perform a volume-weighted average of the biases from the two sets of simulations.

%%%%%%%%%%%%%%%%%%%%%%%%%%%%%%%%%%%%%%%%%%%%%%%%%%%%%%%%%%%%%%%%%%%%%%%%%%%%%
%%%%%%%%%%%%%%%%%%%%%%%%%%%%%%%%%%%%%%%%%%%%%%%%%%%%%%%%%%%%%%%%%%%%%%%%%%%%%
\section{Bias parameters from correlations}
\label{sec:bcorr}

Traditionally bias parameters are used for and measured from $n$-point correlation functions or $n$-spectra. The $n$-th order bias parameters enter the tree-level calculation of the $n+1$-point functions. For instance, $b_1$ appears at the leading order in the large-scale behavior of the halo power spectrum, $b_2$ in the large-scale limit of the bispectrum and $b_3$ in the large-scale limit of the trispectrum. For the comparison to $n$-point functions, we will restrict ourselves to the power spectrum and bispectrum  at tree level here.  The bispectrum also contains nonlocal bias parameters, i.e. biases with respect to the tidal field, that arise from triaxial collapse and gravitational evolution. The estimation of the first and second order bias parameters closely follows the steps outlined in \cite{Baldauf:2012} (see also \cite{Saito:2014}), with the difference that we are performing a joint fit for all the bias parameters, instead of first fitting $b_1$ to the halo power spectrum and then using its value in the bispectrum analysis.

Let us start by discussing the power spectrum. We measure the halo-matter cross power spectrum $P_\text{hm}$, which at tree level (on large scales) is given by
\be
P_\text{hm}(k)=b_1 P_\text{mm}(k).
\ee
We refrain from explicitly including the loop corrections, since they contain third order biases not present in the bispectrum as well as scale-dependent biases $\propto k^2$ \cite{assassi/etal}.  
The advantage of the halo-matter cross power spectrum over the halo-halo power spectrum is that it is free of shot noise. To ensure that our measurements are not contaminated by higher order contributions or scale dependent bias, we will 
in fact fit $P_\text{hm}(k)=(b_1 + b_{P,k^2} k^2 ) P_\text{mm}(k)$ to the simulation
results, where $b_{P,k^2}$ is a free nuisance parameter.  This term absorbs the
loop corrections in the large-scale limit.  
We measure the matter and halo power spectra in the same wavenumber bins in the simulation and take their ratio to cancel the leading cosmic variance, i.e. we define a quantity $q(k)=P_\text{hm}(k)/P_\text{mm}(k)$ and the $\chi^2$
\be
\chi^2_P=\sum_{k}^{k_\text{max}} \left(\frac{q(k)-b_1-b_{P,k^2}k^2}{\sigma[q(k)]}\right)^2,
\ee
where the variance $\sigma^2(q)$ is estimated from the box-to-box scatter between the simulation realizations. 

Let us now turn to the bispectrum. One can form three different bispectra containing the halo field, the halo-halo-halo, the halo-halo-matter and the halo-matter-matter bispectrum. We are using the latter, since it is the only bispectrum free of shot noise. Furthermore, we will employ the unsymmetrized bispectrum, where the halo mode is the one associated with the wavevector $\vec k_3$. This unsymmetrized bispectrum measurement allows for a clear distinction of the second order local bias $b_2$ and tidal tensor bias $b_{s^2}$, once the matter bispectrum is subtracted out. The unsymmetrized tree-level bispectrum reads
\be
B_\text{mmh}(k_1,k_2,k_3)=b_1 B_\text{mmm}(k_1,k_2,k_3) + b_2 P(k_1)P(k_2)+2 b_{s^2} S_2(\vec k_1,\vec k_2) P(k_1)P(k_2)\; ,
\ee
where $B_\text{mmm}$ is the tree-level matter bispectrum (e.g., \cite{Baldauf:2012}), and we employed the tidal operator $S_2$ defined as
\be
S_2(\vec k_1,\vec k_2)=\left(\frac{\vec k_1\cdot \vec k_2}{k_1^2 k_2^2}-\frac{1}{3}\right).
\ee
Similarly to the power spectrum defined above, this bispectrum does not include loop corrections or scale dependent biases. Thus, we again add a term of the form
$b_{B,k^2} (k_1^2+k_2^2)P(k_1) P(k_2)$ with a free coefficient $b_{B,k^2}$,
designed to absorb the loop corrections.  
To cancel cosmic variance, we define the ratio of bispectrum and power spectrum measurements
\be
Q(k_1,k_2,k_3;b_1)=\frac{B_\text{mmh}(k_1,k_2,k_3)-b_1 B_\text{mmm}(k_1,k_2,k_3)}{P_\text{mm}(k_1) P_\text{mm}(k_2)},
\ee
and using this we define the corresponding $\chi^2$
\be
\chi^2_B=\sum_{k_1,k_2,k_3}^{k_\text{max}} \left(\frac{Q(k_1,k_2,k_3;b_1)-b_2  -2b_{s^2}S_2-b_{B,k^2}(k_1^2+k_2^2)}{\sigma[Q(k_1,k_2,k_3;b_{1,\text{fid}})]}\right)^2\; ,
\ee
where the variance of $Q$ is estimated from the box-to-box scatter between the simulation realizations for a fiducial $b_{1,\text{fid}}$. Equivalent results could have been obtained using the estimator presented in \cite{Schmittfull:2014tca}. We decided to stick with the more traditional bispectrum estimation for the following reasons: for their method the smoothing scale of the fields needs to be chosen before the simulation data is reduced, complicating convergence tests. Furthermore, \cite{Schmittfull:2014tca} ignored two-loop corrections to their estimator and higher derivative terms, while we marginalize over an effective shape accounting for the onset of scale dependence. A detailed comparison of the two methods is however beyond the scope of this work.

All measurements are done on the ``lowres'' and ``highres'' sets of the fiducial cosmology.  
We find the best fit biases $b_1$ and $b_2$ by sampling the log-likelihood $\ln \mathcal{L}=-\chi^2_\text{tot}/2$, where $\chi^2_\text{tot}=\chi^2_P+\chi^2_B$ using the Markov Chain code EMCEE \cite{emcee}.  
The errors on the bias parameters are estimated from the posterior distribution of sampling points after marginalizing over the (for our purposes) nuisance parameters $b_{P,k^2}$,  $b_{B,k^2}$ and $b_s^2$.  
We have varied that maximum wavenumber $k_\text{max}$ to ensure that we remain in the regime where the tree level bias parameters remain consistent with increasing $k_\text{max}$.  Further, we demand that the total $\chi^2$ per degree of freedom is approximately unity.  The results shown below use a conservative value of $k_\text{max} = 0.06 \; h \, {\rm Mpc}^{-1}$. This limits the number of modes to $\mathcal{O}(100)$ and thus also the number of power and bispectrum configurations. Due to the cancellation of the leading order cosmic variance this is not of major concern. We have compared the clustering constraints with a larger $2400\; h^{-1}\text{Mpc}$ box providing a factor of 100 more modes to the same cutoff and found consistent results.

%%%%%%%%%%%%%%%%%%%%%%%%%%%%%%%%%%%%%%%%%%%%%%%%%%%%%%%%%%%%%%%%%%%%%%%%%%%%%
%%%%%%%%%%%%%%%%%%%%%%%%%%%%%%%%%%%%%%%%%%%%%%%%%%%%%%%%%%%%%%%%%%%%%%%%%%%%%
\section{Results}
\label{sec:res}

This section presents the results for the Eulerian bias parameters $b_1$ to $b_3$.  For completeness, we also present results for $b_4$, which is poorly constrained, in \refapp{b4}.

In order to obtain a precise comparison between any theoretical prediction for the bias $b_n(M)$ (such as the ESP, \refeq{btheo}) and our data points, we convolve the theoretical prediction with the mass bins used in the simulation (see \refsec{bsep}).  I.e., the theory predictions we will show in the following are given by
\be
b_n^{\rm conv}(M) = \frac{\int{W_n(M',M) n(M') b_n(M') {\rm d} M'}}{\int{ W_n(M',M) n(M'){\rm d}M'}},
\label{eq:b1TbinConv}
\ee 
where $W_n(M',M)$ is the window function of the mass bin given by \refeq{bins}, and $n(M')$ is the differential halo mass function, parametrized by the fitting formula of Eq. (2) in T08.  
In this way, we obtain smooth curves for the theory prediction whose
value at the center of a given mass bin can be compared directly to the simulation results.

%%%%%%%%%%%%%%%%%%%%%%%%%%%%%%%%%%%%%%%%%%%%%%%%%%%%%%%%%%%%%%%%%%%%%%%%%%%%%
\subsection{Linear bias}

\begin{figure}
\centering
\includegraphics[scale=0.55]{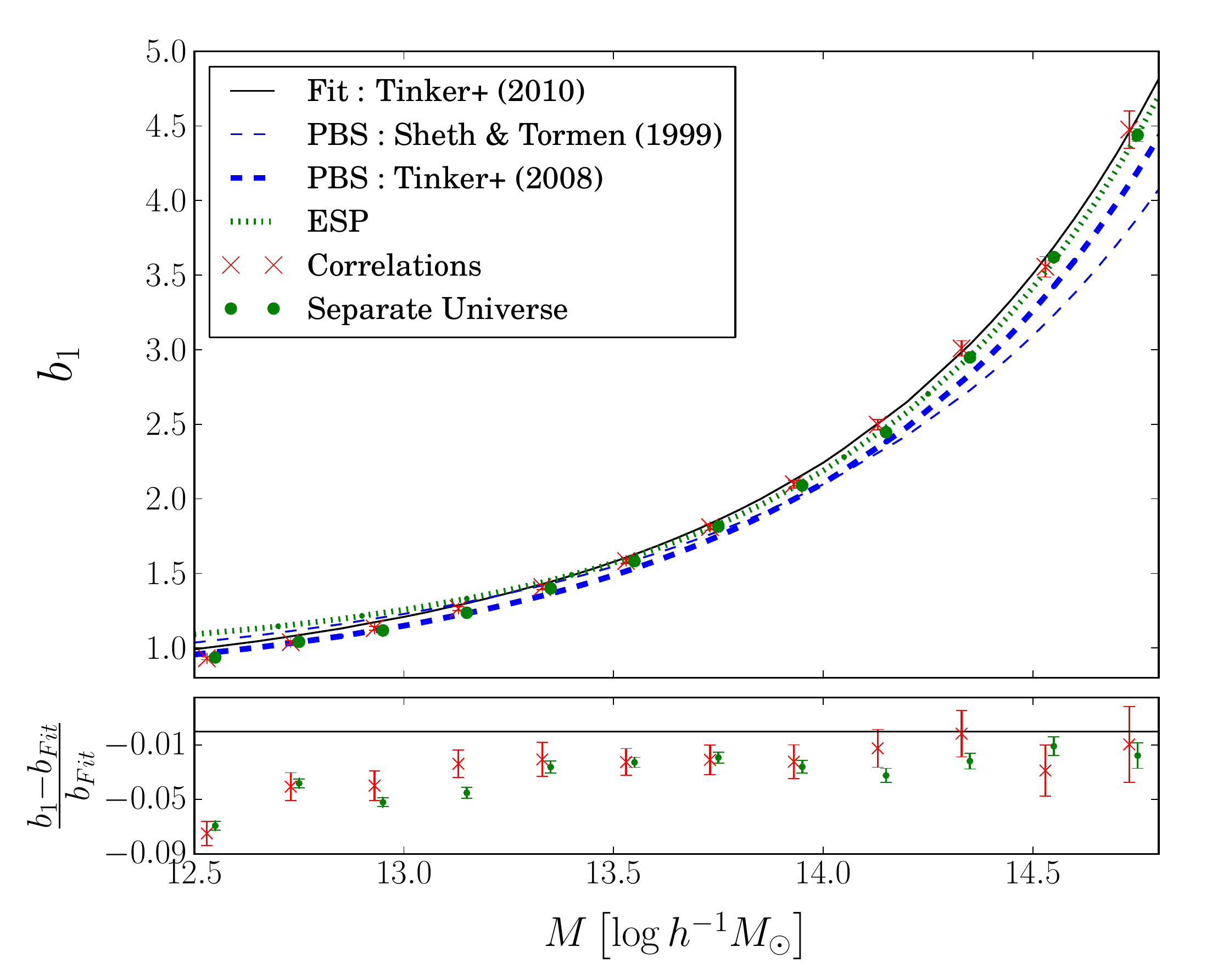} 
\caption{\textbf{Top panel:} comparison between the linear halo bias from separate universe simulations (green dots), and from clustering (red crosses; displaced slightly horizontally for clarity). Error bars that are not visible are within the marker size. The solid black curve is the Tinker et al. (2010) best fit curve for $b_1$, while the dot-dashed green curve is the ESP prediction \refeq{btheo}.  We also show the result obtained by applying the PBS argument [\refeq{biasPBS}] to the T08 
and ST99 mass functions (blue dashed curves). \textbf{Bottom panel:} relative difference between the measurements and the Tinker et al. (2010) best fit.} 
\label{fig:b1}
\end{figure}

\refFig{b1} presents the results for $b_1$. The green points show the results obtained from the separate universe simulations, while the red crosses show those from fitting $P_{\text{hm}}$ and $B_{\text{mmh}}$. The mutual agreement of the two measurements is very good (the only point with relative difference greater than the $1\sigma$ uncertainty is at $\log M=13.15$). The error bars of the separate universe measurements are significantly smaller. Note however that the effective volume used by these measurements is also larger, since the halo-matter power spectrum was only measured in the fiducial boxes. This is a first validation of the separate universe method and also proves its efficiency.

These results are consistent with the ones presented in \cite{li/etal:15} who derived the linear bias from abundance matching.  Since Ref.~\cite{li/etal:15} used a linearized implementation of separate universe simulations, they are restricted to small overdensities (they take $\dr= \pm 0.01$), resulting in very small changes in the halo abundance.  For such small changes, abundance matching is much more efficient than binning halos into finite mass intervals.  We circumvent this issue by using fully nonlinear separate universe simulations which allow us to simulate arbitrary values of $\dr$.  

We also compare our data with several results from the literature. The solid black curve is the fit to $P_{\text{hm}}$ measurements from Tinker et al. (2010) \cite{Tinker:2010} [their Eq. (6)]. As shown in the lower panel of
\refFig{b1}, the agreement is better than 5\%, the quoted accuracy of the
fitting formula. Note that we do not remove unbound particles from our
halos, which we expect to lead to a slight underestimate of the bias at the few percent level at low masses.   
Next, we turn to the ``standard'' peak-background split 
argument \refeq{biasPBS} applied to the universal mass functions of ST99 
and T08 (blue dashed curves).  At low masses, the T08 curve is at 1\% level agreement but the ST99 prediction overestimates the bias by around 8\%.  The agreement is worse at high mass where these two curves underestimate the bias by around 8\%  and 11\% respectively.

The green dot-dashed line finally shows the prediction from excursion set
peaks \refeq{btheo}.  The agreement at high masses is excellent, where
the ESP matches the measured $b_1$ to better than 2\%.  The agreement is far less good at low masses where the ESP prediction overestimates the bias by roughly 10\%.  Note that the assumption that halos correspond to peaks in the
initial density field is not expected to be accurate at low masses
\cite{ludlow/porciani}. Part of the discrepancy might also come from the up-crossing criterion applied to derive the ESP prediction, which is only expected to be accurate at high masses \cite{Musso:2013}. It is worth emphasizing that \refeq{biasPBS} still 
applies in the case of the ESP.  That is, the large-scale bias can still be
derived directly from the mass function.  The key difference to the
PBS curves discussed previously is that, following \cite{Paranjape:2013},
we employ a stochastic moving barrier, which changes the relation between
mass function and bias.  This more realistic barrier leads to the
significant improvement in the prediction of the bias for high-mass halos. 

%%%%%%%%%%%%%%%%%%%%%%%%%%%%%%%%%%%%%%%%%%%%%%%%%%%%%%%%%%%%%%%%%%%%%%%%%%%%%
\subsection{Higher order biases}

\begin{figure}
\centering
\includegraphics[scale=0.5]{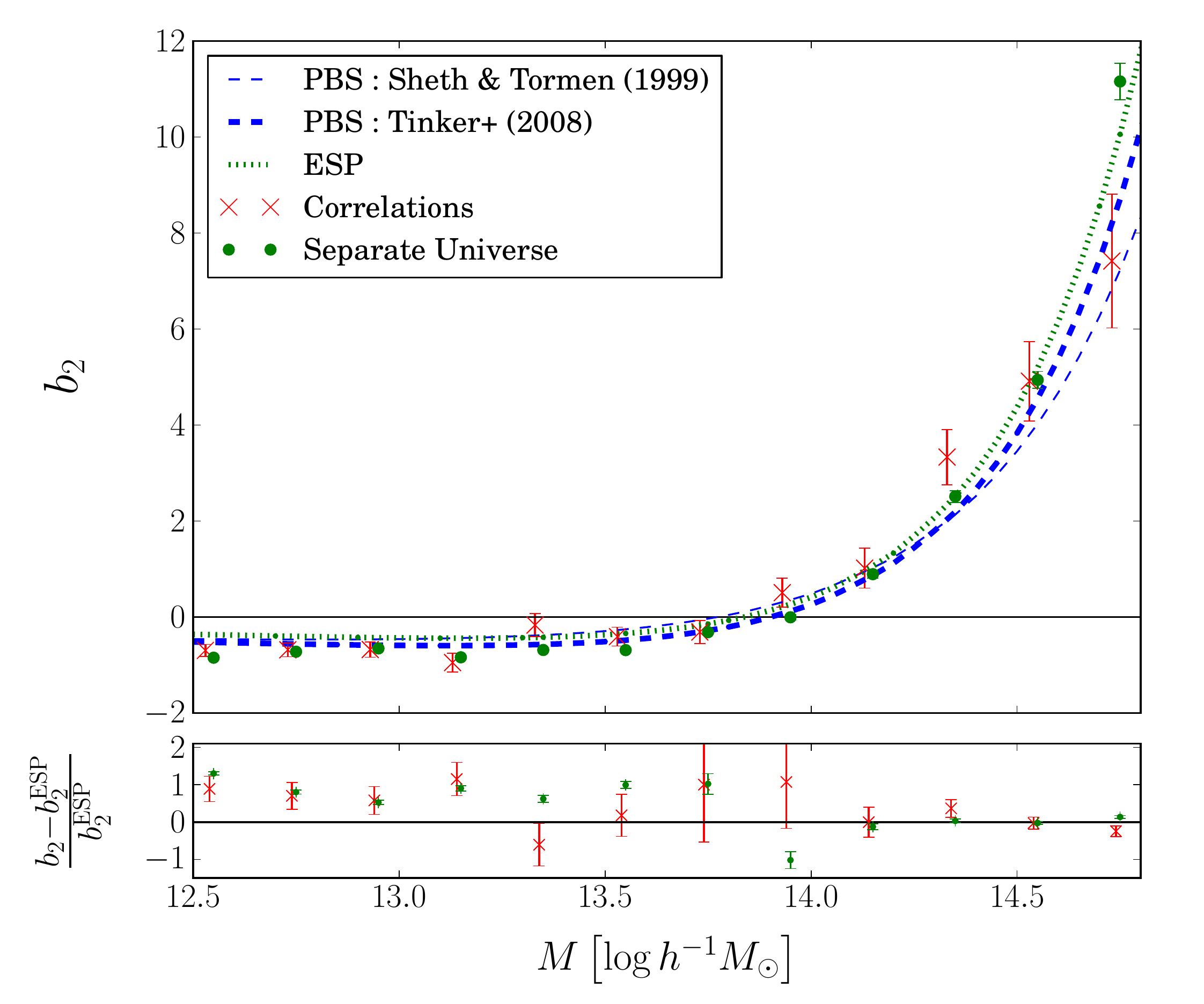} 
\caption{\textbf{Top panel:} same as \refFig{b1}, but for the quadratic bias $b_2$.  The color code is as in \refFig{b1}. \textbf{Bottom panel:} relative difference between measurements and the theoretical prediction of the ESP. In each panel, the clustering points have been horizontally displaced as in \refFig{b1}.} 
\label{fig:b2}
\end{figure}
\begin{figure}
\centering
\includegraphics[scale=0.55]{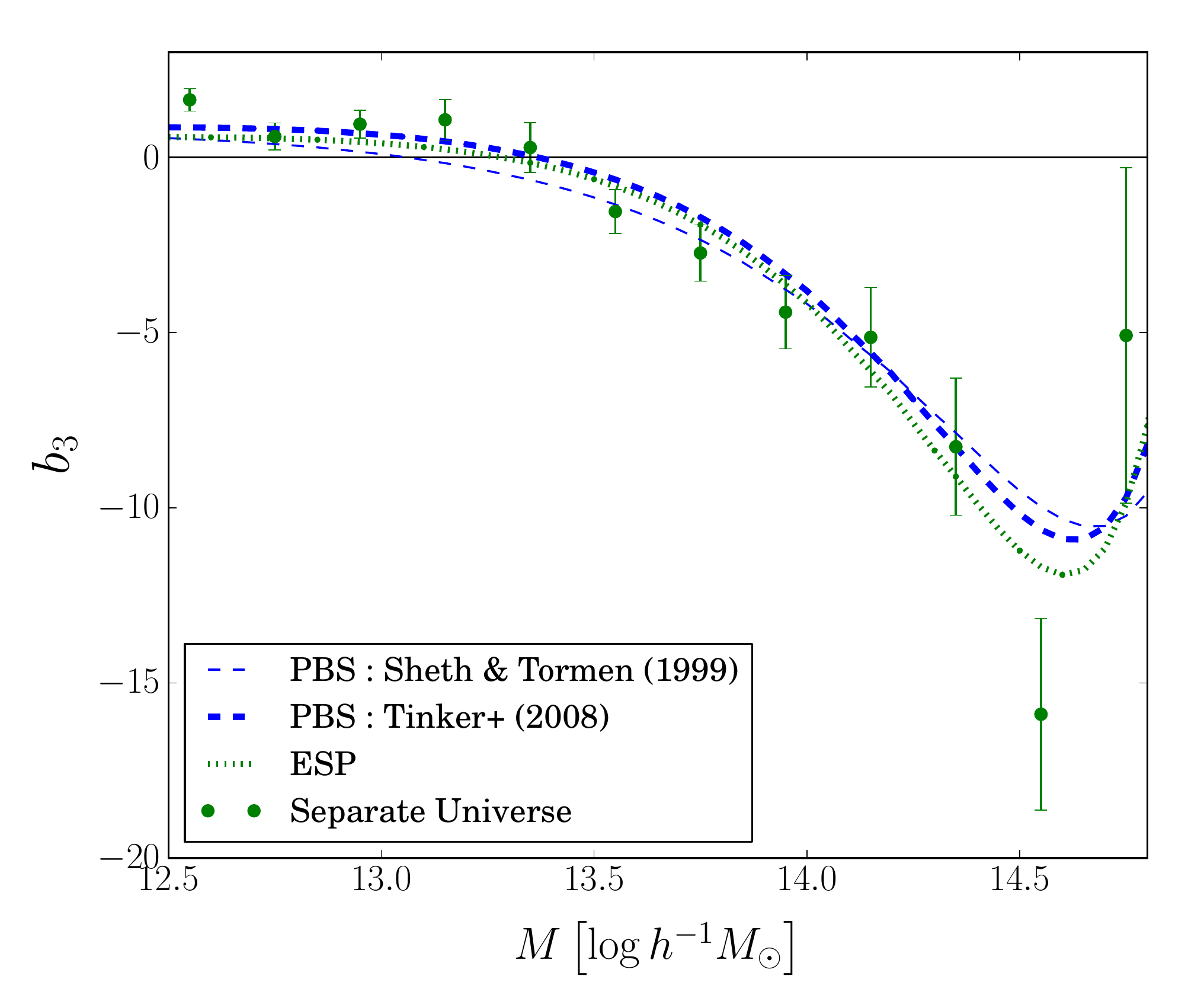} 
\caption{As \refFig{b2} but for $b_3$.} 
\label{fig:b3}
\end{figure}

\refFigs{b2}{b3} present the analogous results of \refFig{b1} for $b_2$ and $b_3$, respectively.  For $b_2$ at masses below $10^{13.5} h^{-1} M_\odot$, there is some scatter in the separate universe results that is apparently larger than what is expected given the error bars (a hint of a similar effect can be seen in $b_1$ as well).  
Note however that there is significant residual degeneracy
between the $b_n$ for a given mass bin, so that a ``$\chi$-by-eye''
can be misleading.  As an example, we show projections of the likelihood for one mass bin in \refFig{contours}.  
The covariance between the bias parameters is further explored in \refapp{cov}.  Covariance in the halo shot noise between different mass bins, which we do not take into account in the likelihood, could also contribute to the fluctuations in the bias parameters.

In the case of $b_2$, we can compare the separate universe results to the results of fitting to $P_{\text{hm}}$ and $B_{\text{mmh}}$.  Again, we find good agreement, with all points being within $2\sigma$ from each other.  Note that $b_2$ is most difficult to constrain from correlations around its zero-crossing.   The difference in constraining power between the two methods is now even larger than in the case of $b_1$.  This is because, when using correlations, $b_2$ has to be measured from a higher order statistic which has lower signal-to-noise.  
In the case of $b_3$, a measurement from correlations would have to
rely on the trispectrum and accurate subtraction of 1-loop contributions in 
perturbation theory.  We defer this significantly more involved measurement
to future work.  
As discussed in the introduction, it is difficult to rigorously compare these
measurements to previously published results, since those were measured
at a fixed smoothing scale and did not take into account nonlocal bias
terms.  Nevertheless, our results for $b_2$ and $b_3$ appear broadly consistent with those of 
\cite{angulo/baugh/lacey:2008,paranjape/etal:2013}
and \cite{angulo/baugh/lacey:2008}, respectively.

We again compare with the peak-background split results, now derived at
second and third order from the ST99 and T08
mass functions.  For $b_2$, at low mass, both predictions deviate from our measurements by about 50\%. At high mass, the deviation is at most 25\% for T08 and 40\% for ST99. In the low mass range, this apparently big discrepancy is also due to the smallness of the absolute value of $b_2$. 
In the case of $b_3$, the PBS predictions using either the T08 or ST99
mass functions are in fact completely consistent with the measurements 
at masses $\gtrsim 10^{12.7} h^{-1} M_\odot$ and $10^{13.5} h^{-1} M_\odot$, respectively.

Turning to the ESP prediction, we again find very good agreement at
high masses, although for $b_2$ and $b_3$ the performance is not significantly better than the
PBS-derived biases from the T08 mass function.  At low masses,
we again find larger discrepancies, with the ESP now underpredicting
the magnitude of $b_2$ and $b_3$.  The same caveats regarding the relation of low-mass halos to peaks and the efficiency of the up-crossing condition apply here, i.e. we do not expect the ESP
prediction to work well for those masses.\\  

\begin{figure}
\centering
\includegraphics[scale=0.37]{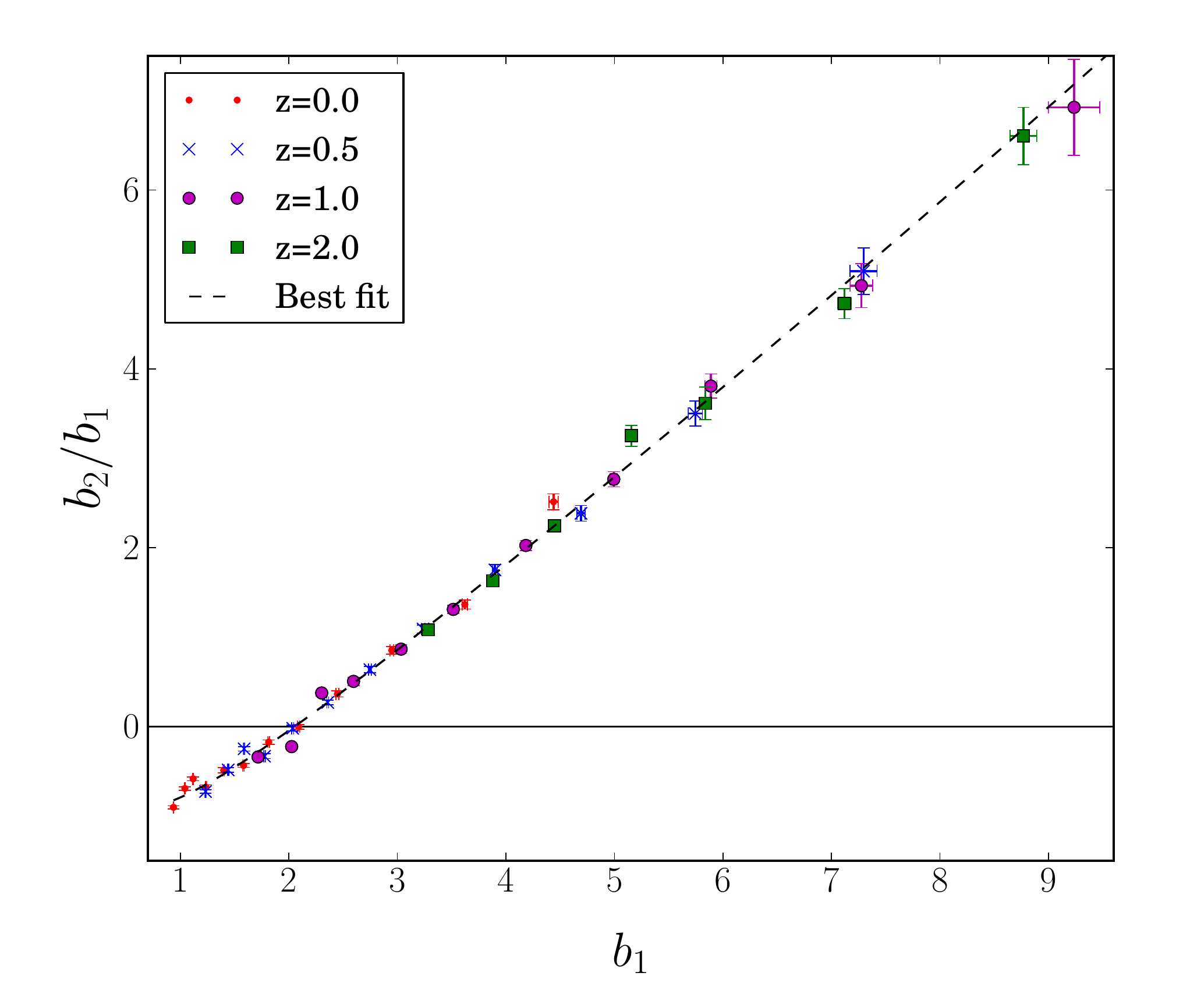} 
\includegraphics[scale=0.37]{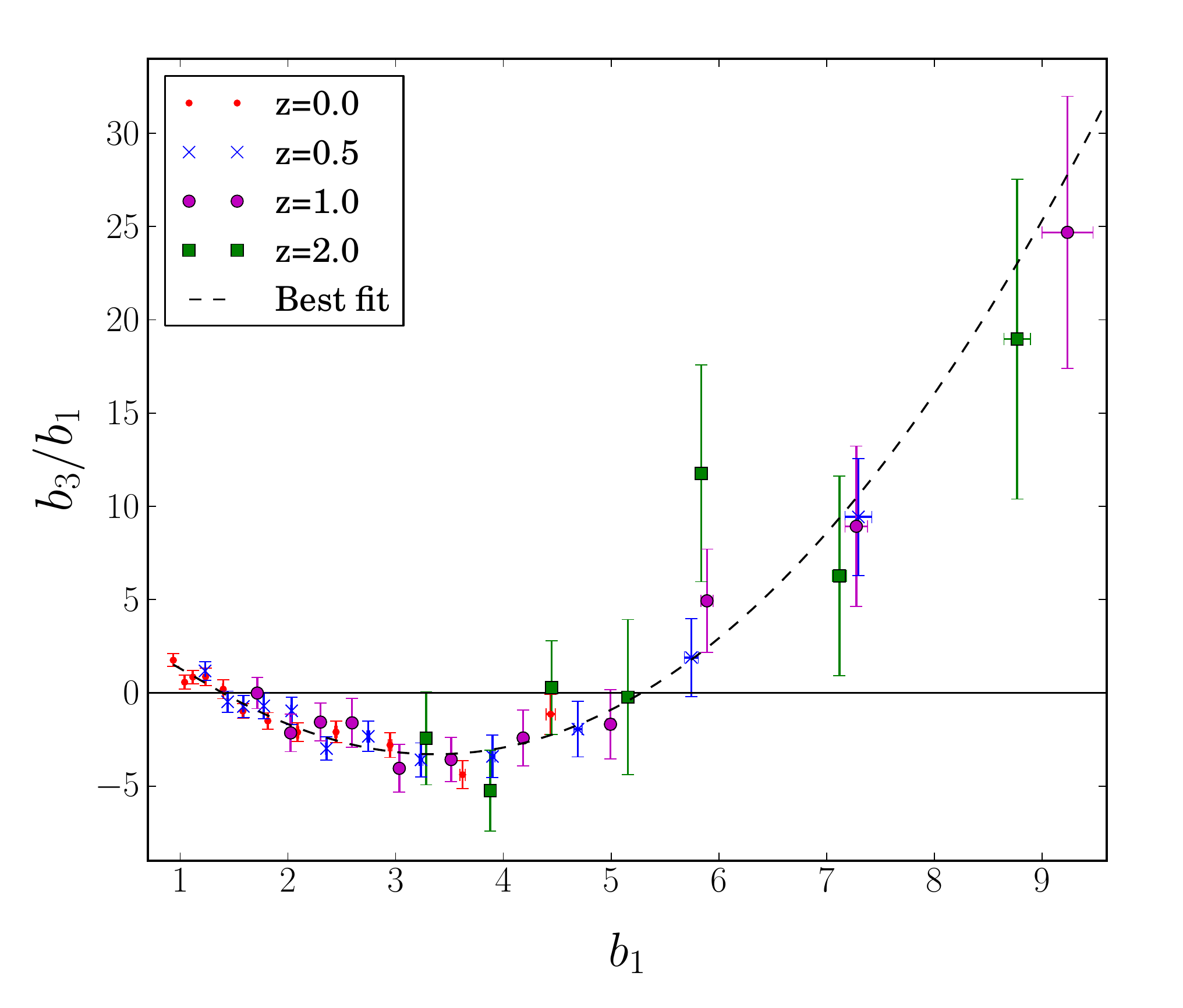}
\caption{$b_2$ and $b_3$ as a function of $b_1$ obtained from separate universe simulations and for different redshifts. The dashed curves present the third order best fit polynomial for each bias. See text for details about the fit.} 
\label{fig:b2b3(b1)}
\end{figure}

So far, we have only shown results at redshift $0$.  \refFig{b2b3(b1)}
shows results from various redshifts by plotting $b_2,\,b_3$ as functions of
$b_1$.  If the bias parameters are uniquely determined by $\sigma_0 = \sigma(M)$, then this relation will be redshift-independent.  Indeed, we find no
evidence for a redshift dependence over the range $z = 0 \dots 2$ and
$b_1 = 1 \dots 10$.  Note that we have kept the overdensity criterion
$\Delta_{\rm SO}=200$ fixed.  
Since the separate universe simulation measurements of $b_2$ and $b_3$ 
are very accurate, we provide fitting formulas in 
the form of $b_n(b_1)$ for convenience. 
Given the consistency with a universal behavior, we perform a joint fit of results from all redshifts. We use a $3^{\rm rd}$ order polynomial form for both $b_2$ and $b_3$. Again, we use a weighted least squares method for the fit but do not take into account the error on $b_1$ since it is much smaller than those in $b_2,\,b_3$. We obtain
\be
b_2(b_1) = 0.412-2.143\,b_1+0.929\,b_1^2+0.008\,b_1^3, 
\label{eq:b2(b1)} 
\ee
and
\be
b_3(b_1) = -1.028+7.646\,b_1-6.227\,b_1^2+0.912\,b_1^3. 
\label{eq:b3(b1)} 
\ee
The fits are shown as dashed lines in the two panels of \refFig{b2b3(b1)}. Notice that we restricted ourselves to $b_1 < 9.6$ on these figures for clarity but we used the full range of results to produce the fits.
Note that one should be careful when using these formulas outside the fitting range $1\lesssim b_1\lesssim 10$.
\refeqs{b2(b1)}{b3(b1)} are similar to the fitting formulas provided in \cite{Hoffmann:2015} who fitted $2^{\rm nd}$ and $3^{\rm rd}$ order polynomials for
$b_2(b_1)$ and $b_3(b_1)$, respectively, to PBS predictions, and found no redshift dependence of their results. Such universal relations became already apparent in \cite{Saito:2014} (their figure 9). 

%%%%%%%%%%%%%%%%%%%%%%%%%%%%%%%%%%%%%%%%%%%%%%%%%%%%%%%%%%%%%%%%%%%%%%%%%%%%%
%%%%%%%%%%%%%%%%%%%%%%%%%%%%%%%%%%%%%%%%%%%%%%%%%%%%%%%%%%%%%%%%%%%%%%%%%%%%%
\section{Conclusions}
\label{sec:concl}

We have presented a new method to measure the large-scale, renormalized
local density bias parameters $b_n$ of dark matter halos, with $n=1,2,3$,
by running simulations which simulate an infinite-wavelength density
perturbation of arbitrary amplitude.  This method can be seen as an 
exact implementation of the peak-background split.  
This method has several advantages, including a simple implementation applicable,
in principle, to arbitrarily high $n$.  The most important advantage,
however, is that the measured biases are not affected
by the modeling of scale-dependent or nonlinear corrections, and there
is no ambiguous choice of $k_{\rm max}$, with the associated risk of 
overfitting, as when
fitting halo $N$-point functions.  The most significant disadvantage
of the method is that it needs a set of dedicated simulations with
varying cosmological parameters to generate a range of $\d_L$ (note
however that once the simulations are done, they can be used for
various studies, such as for example the nonlinear power spectrum
response \cite{Wagner:2015}).  

We have compared our results for $b_1$ and $b_2$ to those measured
from the halo-matter power spectrum and halo-matter-matter bispectrum,
and find excellent agreement overall.  One necessary condition for this
agreement is a careful fitting procedure of the halo statistics and choice of $k_{\rm max}$.  

We also compared our results to predictions based on the analytical peak-background
split.  Once a specific barrier $B$ is assumed, the PBS allows for
a derivation of all local bias parameters $b_n$ from a given halo mass
function.  The simplest and most common choice is $B=\d_c$, which we have
applied to the ST99 and T08 mass function prescriptions.  
We found that even though the latter provides a very accurate mass function, the linear bias derived via the PBS and simple collapse threshold is only accurate at the $\sim 10$\% level, in agreement
with previous results \cite{manera/etal:2010}.  Things are even worse for $b_2$, with up to 50\% discrepancy at low mass, although the absolute difference between the PBS predictions and the measurements is similar to that in $b_1$.  For $b_3$, the simple PBS predictions are consistent with the measurements (at least at high masses), but this is not a very strong statement given the large error bars on $b_3$.  

We also derived the biases predicted in the excursion set-peaks
approach, which includes a stochastic moving barrier motivated by 
simulation results.  At high mass, this performs much better, at least for $b_1$, showing
that the choice of barrier is a key ingredient in deriving accurate
bias parameters.  In this context, it is important to note that previous
results on the inaccuracy of PBS bias parameters \cite{manera/etal:2010}
relied on the simple constant threshold $B=\d_c$.  This shows that the cause of 
theses inaccuracies is not the peak-background split itself.  The
inaccuracy of the peak-background split thus depends on what one
defines PBS to mean, and can be summarized as follows:
\begin{itemize}
\item The PBS implemented via the separate universe approach is exact.
\item The PBS using a simulation-derived stochastic moving barrier \cite{robertson/etal,Paranjape:2013}, as in the ESP, is accurate to a few percent, at least at high masses.  The discrepancy found at low mass can be explained by the failure of the peak assumption at such masses, an issue unrelated to the choice of the barrier.
\item The PBS using the constant spherical collapse barrier is no better than 10\%\,.
\end{itemize}  

We also provide fitting formulas for $b_2,\,b_3$ as a function of $b_1$
which are valid over a range of redshifts and 
can be useful for predictions and forecasts based on halo statistics,
such as for example the halo model.  

In the future, we plan to extend our analysis to accurately measure
assembly bias, i.e. the dependence of bias on halo properties beyond
the mass (e.g., \cite{gao/etal,wechsler/etal,dalal/etal}).  
Further, it will be interesting to extend this technique beyond the 
infinite wavelength, spherically symmetric ``separate universe'' to allow for precision measurements of the 
tidal and scale-dependent biases. 

\acknowledgments{We thank Aseem Paranjape, Marcello Musso and Vincent~Desjacques for useful discussion about the ESP. F.S.~acknowledges support from the Marie Curie Career Integration Grant  (FP7-PEOPLE-2013-CIG) ``FundPhysicsAndLSS''. T.B.~gratefully acknowledges support from the Institute for Advanced Study through a Corning Glass works foundation grant.}

%%%%%%%%%%%%%%%%%%%%%%%%%%%%%%%%%%%%%%%%%%%%%%%%%%%%%%%%%%%%%%%%%%%%%%%%%%%%%
%%%%%%%%%%%%%%%%%%%%%%%%%%%%%%%%%%%%%%%%%%%%%%%%%%%%%%%%%%%%%%%%%%%%%%%%%%%%%
\appendix

\section{Halo bias from excursion set peaks}
\label{app:ESP}

In this appendix, we present details of the derivation of the Lagrangian bias parameters of halos in the ESP formalism.  We first introduce some notation, following  \citep{Paranjape:2013}. 
The top-hat and Gaussian filters in Fourier space are given by
\ba
W_\T(kR_\T) & =  \frac{3}{(kR_\T)^3}\left[{\rm sin}(kR_\T)-kR_\T {\rm cos}(kR_\T)\right], \label{eq:WTH} \\
W_\G(kR_\G) &= {\rm e}^{-(kR_\G)^2/2}, \label{eq:WG}
\ea
respectively. The Gaussian filtered spatial moments are defined as
\be
\s{j, \G}^2 \equiv \int {\rm dln}k \D^2 (k) k^{2j} W_\G(kR_\G)^2, \hspace{0.2cm}  j\geq 1,
\label{eq:sigmaG}
\ee
and the first mixed moment as
\be
\s{1,{\rm m}}^2 \equiv \int {\rm dln}k \D^2 (k) k^{2} W_\G(kR_\G)W_\T(kR_\T).
\label{eq:sigma1m}
\ee
From these quantities we build the characteristic length 
\be
R_\ast \equiv \sqrt{3}\frac{\s{1,\G}}{\s{2,\G}}, 
\label{eq:Rstar}
\ee
and the spectral moment
\be
\gamma \equiv \frac{\s{1,{\rm m}}^2}{\s{0}\s{2,\G}}.
\label{eq:gamma}
\ee

Let us begin with Gaussian peaks.  Ref.~\citep{Bardeen:1986} showed that the density of peaks of scaled height $\nu$ of a Gaussian-filtered Gaussian random field is
\be 
n_{\rm pk}(\nu) = \int {\rm d}x\, n_{\rm pk}(\nu,x) = \frac{{\rm e}^{-\nu^2/2}}{\sqrt{2\pi}}\frac{G_0(\gamma,\gamma\nu)}{(2\pi R_\ast ^2)^{3/2}},
\label{eq:npk}
\ee 
where $x=-\nabla^2 \d / \sigma_2$ is related to the curvature of the field and   
\be 
G_\alpha(\gamma ,x_\ast)\equiv \int {\rm d}x\, x^\alpha F(x) p_{\rm G}(x-x_\ast ; 1-\gamma^2),
\label{eq:G}
\ee
where $ p_{\rm G}(x-\mu ; \sigma^2)$ is a Gaussian distribution with mean $\mu$ and variance $\sigma^2$ and $F(x)$ is the peak curvature function (Eq. (A15) of BBKS). Notice that in the particular case of a Gaussian filter $R_\G \nabla^2 \d= \partial \d / \partial R_{\rm G}$ so that $x$ is associated with the curvature of the density field as well as its derivative with respect to the smoothing scale.  In the original peak model \cite{Bardeen:1986}, the smoothing scale $R$ as well as threshold $B$ are fixed, and the peak density is a local function of the smoothed density field $\nu$ and its derivatives.   
On the other hand, in the excursion set, $\nu$ is defined at a fixed location, and varies as a function of the smoothing scale.  Hence one must be careful when combining the two. 

We now apply the excursion set argument on the peaks: on a given smoothing scale $\s{0}$ we consider only the peaks that have a smaller height at an infinitesimally larger smoothing scale.\footnote{This condition is actually not exactly the excursion set condition as expressed by \citep{Bond:1991}. Indeed we should ask that the peak height be smaller on every smoothing scale larger than $\s{0}$ but this condition is hard to implement. Ref.~\citep{Musso:2012} showed that the much simpler condition that we use here (involving only a single infinitesimally larger smoothing scale), also called up-crossing condition, already gives very accurate predictions.} We start with the case of a constant barrier of height $\d_c$. In this case, we ask that the scaled peak height lies between $\nu = \d_c/\s{0}$ and $\nu + ({\rm d}\nu/{\rm d}\s{0})\D \nu$. This leads to 
\be 
n_{\rm ESP}(\nu) =\frac{1}{\gamma\nu}\int_0^\infty {\rm d}x\, x n_{\rm pk}(\nu,x),
\label{eq:nesp}
\ee
so that the fraction of mass in peaks of height $\nu$ is
\be
f_{\rm ESP}(\nu)= Vn_{\rm ESP}(\nu) = \frac{{\rm e}^{-\nu^2/2}}{\sqrt{2\pi}}\frac{V}{V_\ast}\frac{G_1(\gamma,\gamma\nu)}{\gamma\nu},
\label{eq:fesp}
\ee
where $V_\ast = (2\pi R_\ast)^{3/2}$ and $V$ is the Lagrangian volume associated to the peak and depends on the filter's shape ($V=4\pi R^3_{\rm TH}/3$ for a top-hat filter).

We follow \citep{Paranjape:2013}, who chose a top-hat filter and improved the model by adding a mass-dependent stochastic scatter to the threshold [see \refeq{barrier}].  
The peaks on the other hand are defined using a Gaussian filter (which ensures that the higher moments $\sigma_j^2,\,j>0$ exist).  We thus need a mapping between the Gaussian scale $R_{\rm G}$ and the top-hat one $R_{\rm TH}$ to ensure that the peaks identified in the Gaussian-filtered density field have density contrast $\d_{\rm TH} = \d_c$ when smoothed with a top-hat filter. Following \citep{Paranjape:2013}, we do that by requiring $\<\d_{\rm TH}\d_{\rm G}\>= \<\d_{\rm TH}^2\>$ which leads to $ R_{\rm G}=0.46 R_{\rm TH}$ with a mild mass dependence that we will not account for.

Including the stochastic parameter $\beta$, the fraction of mass corresponding to \refeq{fesp} is now given by
\be
f_{\rm ESP}(\nu) = \int {\rm d}\beta\, f_{\rm ESP}(\nu | \beta) p(\beta),
\label{eq:fespb}
\ee
where $f_{\rm ESP}(\nu | \beta)$ is the mass fraction at fixed $\beta$ and is given by (Eq. (14) of \cite{Paranjape:2013})
\be 
f_{\rm ESP}(\nu | \beta) = \frac{{\rm e}^{-(\nu+\beta)^2/2}}{\sqrt{2\pi}}\frac{V}{V_\ast} \int_{\beta\gamma}^\infty {\rm d}x \frac{x-\beta\gamma}{\gamma\nu}F(x)p_{\rm G}(x-\beta\gamma-\gamma\nu;1-\gamma^2),
\label{eq:fespcond}
\ee
where $V=4/3\pi R_{\rm TH}^3$ is the volume associated with a top-hat filter. Given a probability distribution function (PDF) for $\beta$ we can thus compute $f_{\rm ESP}(\nu)$ with \refeq{fespb} and, applying Bayes' theorem, we can compute the PDF for $\beta$ at fixed $\nu$
\be 
p(\beta|\nu) = \frac{f_{\rm ESP}(\nu | \beta) p(\beta)}{f_{\rm ESP}(\nu)},
\label{eq:pofbnu}
\ee
which we will need to compute the Lagrangian bias parameters. 

We can now give predictions for the Lagrangian halo bias, inserting \refeq{fespcond} for the multiplicity function into \refeq{nf}.  We then apply the PBS argument as described in \refsec{PBS} and send $\nu = \d_c/\s{0}$ to $\nu_1 = \nu\left(1-\d_0/\d_c\right)$.  Notice that the stochastic part of the barrier \refeq{barrier} is not modified.  Further, the shift in the barrier (and hence in $\nu$) should not be applied to the denominator of \refeq{fespcond} as this factor of $\nu$ only appears when one changes variables from $sf(s)$ to $\nu f(\nu)$ and is physically unrelated to the barrier.  We then use \refeq{biasPBS} to find the bias parameters at fixed $\beta$. To obtain the large-scale Lagrangian bias as measured in simulations, one must further marginalize over $\beta$. This finally yields
\be
\d_c^n b_n(\nu) = \sum_{i=0}^n \binom{n}{i} \int {\rm d}\beta\, p(\beta|\nu) \mu_i(\nu,\beta)\lambda_{n-i}(\nu,\beta), 
\label{eq:btheo}
\ee
with $p(\beta|\nu)$ given by \refeq{pofbnu} and 
\ba
\mu_n (\nu,\beta)& = \nu^n H_n(\nu+\beta),\vs
\lambda_n(\nu,\beta) & = (-\Gamma\nu)^n\<H_n(y-\beta\Gamma-\Gamma\nu)|\nu,\beta\>_y,
\label{eq:mulambda}
\ea
where $H_n$ is the $n^{\rm th}$ order Hermite polynomial, $\Gamma \equiv \gamma/\sqrt{1-\gamma^2}$ and we defined,
for any function $h(y,\nu,\beta)$,
\ba
\<h(y,\nu,\beta)|\nu,\beta\>_y \equiv \frac{\int_{\beta\Gamma}^\infty {\rm d}y(y-\beta\Gamma)F(y\gamma/\Gamma)p_{\rm G}(y-\beta\Gamma-\Gamma\nu ;1)h(y,\nu,\beta)}{\int_{\beta\Gamma}^\infty {\rm d}y(y-\beta\Gamma)F(y\gamma/\Gamma)p_{\rm G}(y-\beta\Gamma-\Gamma\nu ;1)}.
\label{eq:moyh}
\ea
\refeq{btheo} gives the theoretical predictions we compare our results with. This result is the same as the one given in \cite{Paranjape:2012,Paranjape:2013} who used 
\be
\<1+\d_h | \d_0, S_0\> \equiv \frac{f(\nu | \d_0,S_0)}{f(\nu)} = \sum_{n=0}^\infty \frac{\d_0^n b_n}{n!},
\label{eq:dh}
\ee 
as the definition of the bias parameters \cite{Mo:1996,Mo:1997}, which defines the overdensity of halos $\d_h$ and emphasizes the fact that, in the ESP formalism the effect of the underlying dark matter density field on the abundance of halos of mass $M$ can be estimated from the conditional fraction $f(\nu | \d_0,S_0)$ of walks that first crossed the barrier of height $\nu$ on scale $\s{0}$ having passed through $\d_0 < B$ on scale $S_0 < \s{0}$ before.

\section{Comparison of Lagrangian and Eulerian separate universe biases}
\label{app:compbsep}

\begin{figure}
\centering
\includegraphics[scale=0.5]{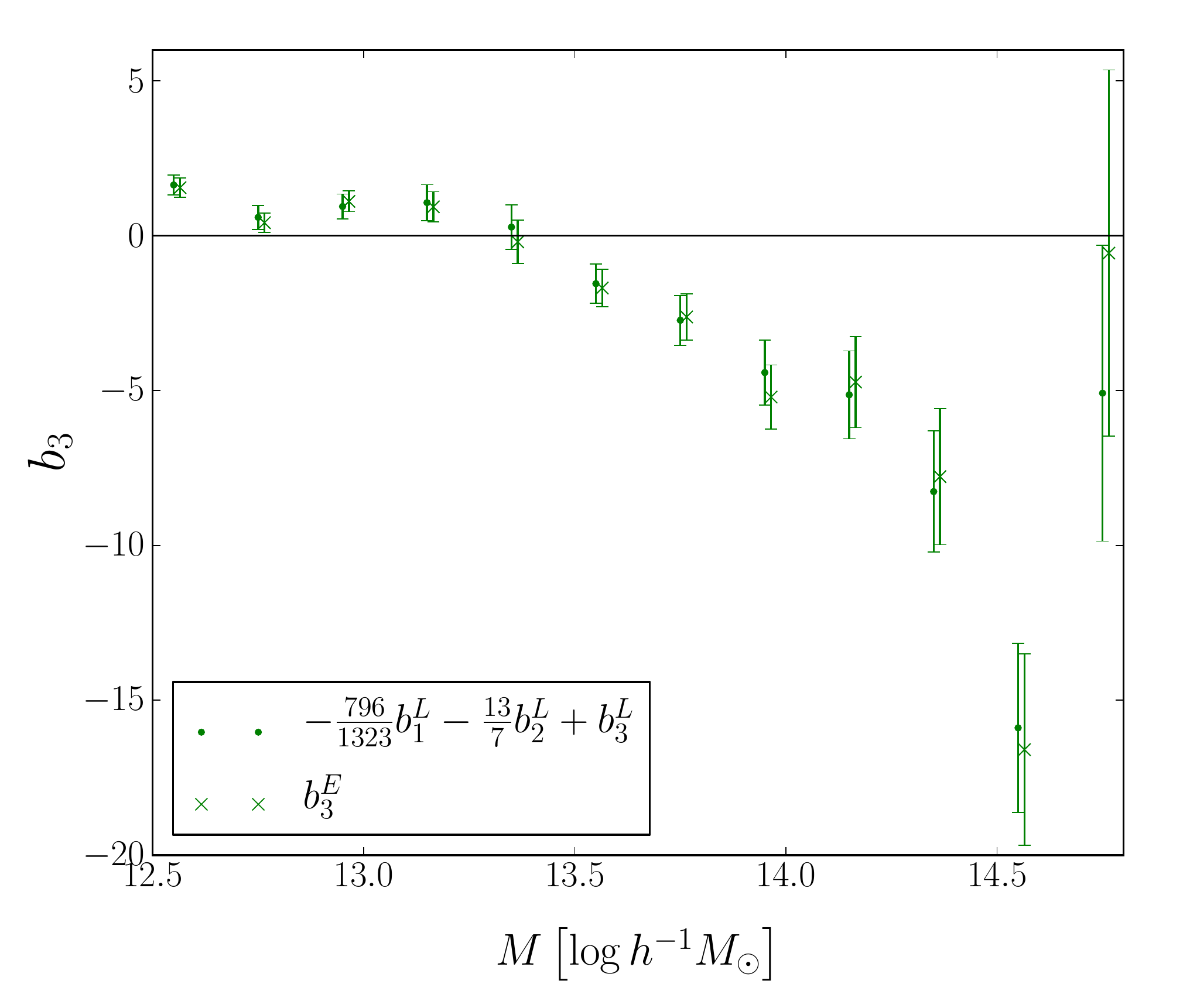}
\caption{Comparison of $b_3$ measured directly via fitting $\d_h$ vs the Eulerian density $\d_\rho$ (crosses) and, the corresponding value inferred from the Lagrangian bias fits, as shown in the main text (dots). The crosses have been displaced horizontally for clarity.} 
\label{fig:b3_E_from_L}
\end{figure}

To derive the Eulerian bias parameters from the Lagrangian ones, we use the spherical collapse model (which is exact in our case). To do that, we use the result (B.18) of \citep{Wagner:2015} linking $\dr(t)$ to $\d_L$. Setting $t=t_0$ (present time) yields $a(t_0)=1$ which leaves us with
\be
\dr = \sum_{n=1}^{\infty}f_n\d_L^n,
\label{eq:dr_dL}
\ee
with $f_n$ being constant coefficients given in their appendix B.  Note that these numbers are derived for a flat matter-dominated (Einstein-de Sitter) universe.  However, ref.~\cite{Wagner:2015} checked that they are also accurate at the sub-percent level for $\Lambda$CDM so that the difference to the exact coefficients for $\Lambda$CDM is completely negligible compared to the uncertainties of the measured bias parameters.  Then, using the continuity equation for the dark matter density as well as for the density of dark matter halos, the fact that the two comove on large scales, and neglecting $\dr$ at very early times, we find
\be
1+\d_h = (1+\dr)\times(1+\d_h^L).
\label{eq:dhE_dhL}
\ee
Finally, we have the bias relations
\be
\d_h^L = \sum_{n=1}^{\infty}\frac{b_n^L}{n!}\d_L^n, \quad
\d_h = \sum_{n=1}^{\infty}\frac{b_n}{n!}\dr^n.
\label{eq:BiasExp_E}
\ee
Plugging \refeq{dr_dL} and \refeq{BiasExp_E} into \refeq{dhE_dhL}, we find 
\be
b_1 = 1+b_1^L,
\label{eq:b1_E}
\ee
\be
b_2 = \frac{8}{21}b_1^L+b_2^L,
\label{eq:b2_E}
\ee
\be
b_3 = -\frac{796}{1323}b_1^L-\frac{13}{7}b_2^L+b_3^L,
\label{eq:b3_E}
\ee
\be
b_4 = \frac{476320}{305613}b_1^L+\frac{7220}{1323}b_2^L-\frac{40}{7}b_3^L+b_4^L .
\label{eq:b4_E}
\ee
We can thus compare the Eulerian bias parameters determined from the
measured Lagrangian parameters using these relations with the direct
Eulerian measurement.  This result is shown for $b_3$ in \refFig{b3_E_from_L} for a $5^{\rm th}$ order fit.  Clearly, the bias parameters agree very well.  
This also holds for $b_1,\,b_2$ even though we do not show it here.  
We have found, however, that the polynomial fit is slightly more stable
when measuring the Lagrangian bias parameters, i.e. fitting to $\delta_L$ rather than $\dr$.  
In particular, the covariance between the Eulerian bias parameters $b_n$ and $b_{n+1}$ is reduced when they are derived from the measured Lagrangian bias parameters instead of measuring them directly.
This is because the simulated positive and negative values for
$\d_L$ are almost symmetric, whereas those for $\dr$ are not due to nonlinear evolution.

\section{Effect of the degree of the fitting polynomial on the bias parameters}
\label{app:deg}

\begin{figure}
\centering
\includegraphics[scale=0.5]{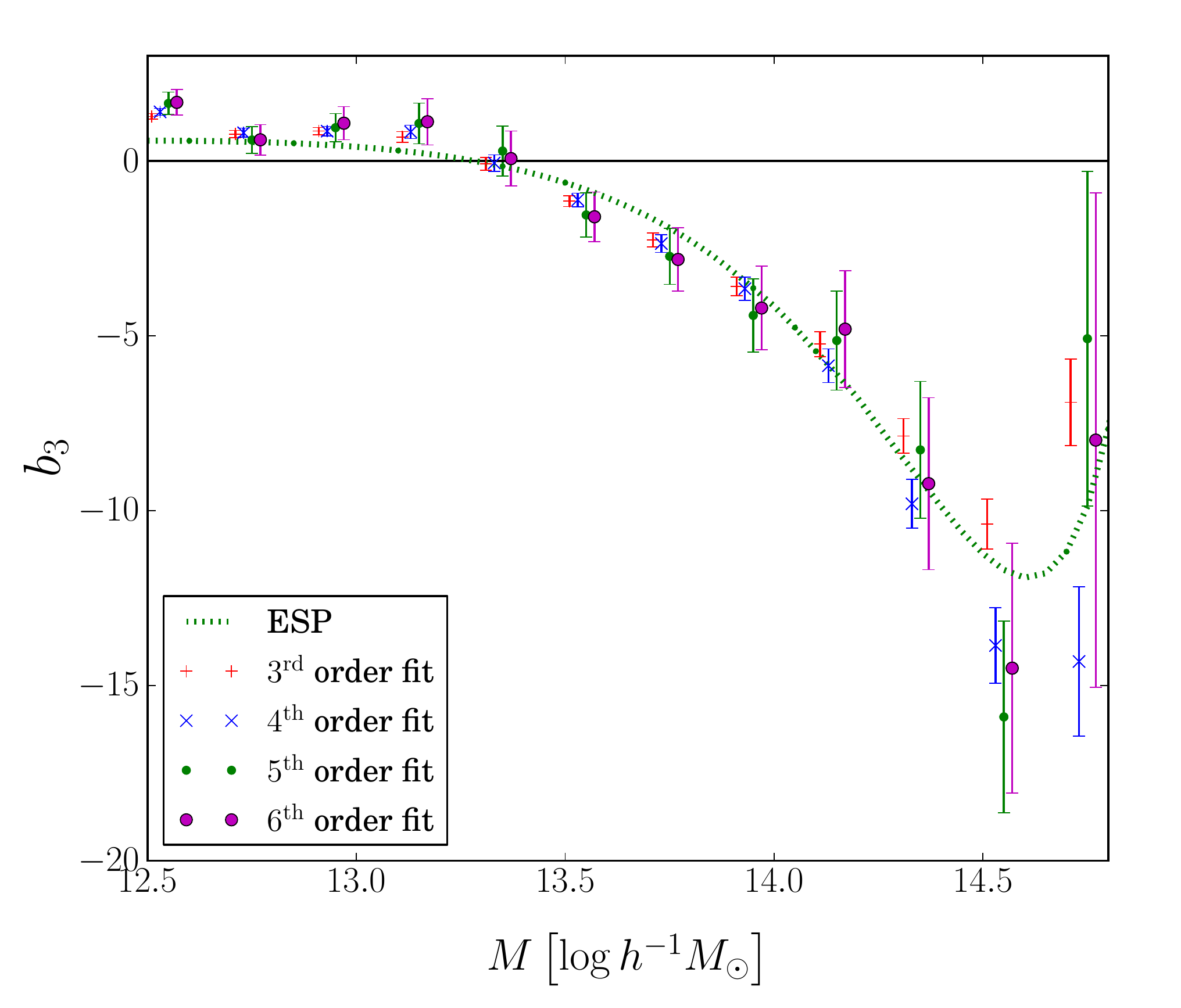} 
\caption{Eulerian $b_3$ obtained with four different fitting polynomial orders from 3 to 6.  For reference, we also show the ESP prediction.} 
\label{fig:b3deg}
\end{figure}

We study here how the degree of the fitting polynomial affects the results.  
This, along with the covariance between the bias parameters presented in \refapp{cov},
will justify our choice of using a $5^{\rm th}$ order fit for $b_1$ to $b_3$ and a $6^{\rm th}$ order one for $b_4$.  We present results for $b_3$ as this bias parameter is the most sensitive to the degree of the fit, but the effect is qualitatively the same for every bias parameter.  \refFig{b3deg} shows the results obtained with four different degrees from $3$ to $6$. 

Setting the degree of the fitting polynomial is a balance between a bias 
in resulting fit parameters (when reducing the polynomial order) and 
increasing measurement errors (when increasing the order).  
As \refFig{b3deg} shows, we need to go to a $5^{\rm th}$ order fit in order to have convergence for $b_3$ (i.e. so that the results obtained with the $n^{\rm th}$ order fit are within the error bars of those obtained with a $n+1$ order polynomial).  Hence, to ensure that we obtain unbiased results, we use a $5^{\rm th}$ order fit to obtain $b_1$ to $b_3$ and a $6^{\rm th}$ order fit for $b_4$.

\section{Fourth order bias}
\label{app:b4}

\begin{figure}
\centering
\includegraphics[scale=0.55]{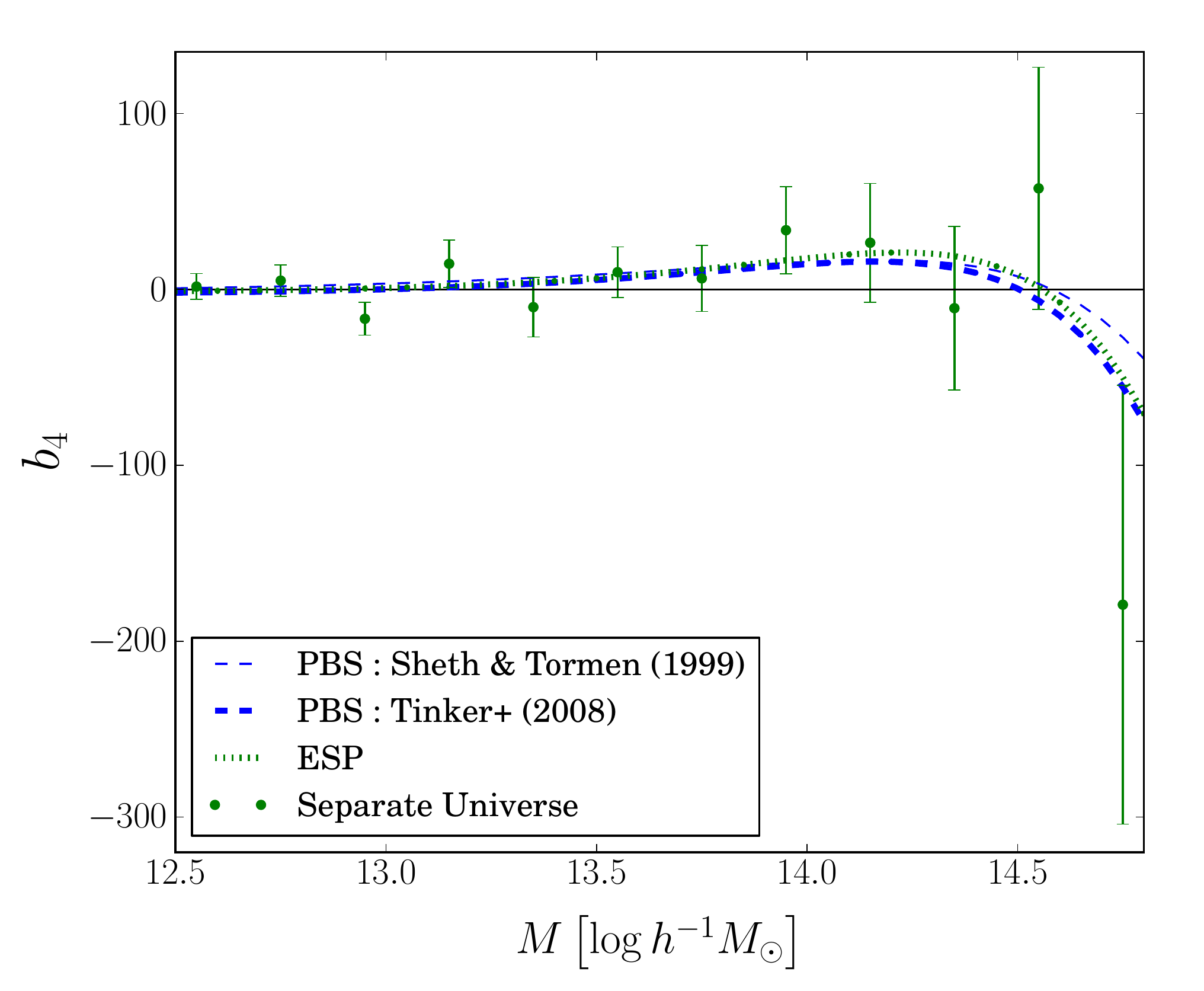} 
\caption{Separate universe results for $b_4$.  The curves are as in \reffig{b2}.} 
\label{fig:b4}
\end{figure}

We present here measurements of $b_4$ that were obtained with a $6^{\rm th}$ order polynomial fit (\refFig{b4}). The scatter in the points and the uncertainties are quite large for this bias parameter, and only an indication of the general behavior can be seen.  For this reason we do not show it in the main text. Nevertheless, the ESP and standard PBS predictions are consistent with the measurements.

\section{Covariance between the bias parameters}
\label{app:cov}

\begin{figure}
\centering
\includegraphics[scale=0.35]{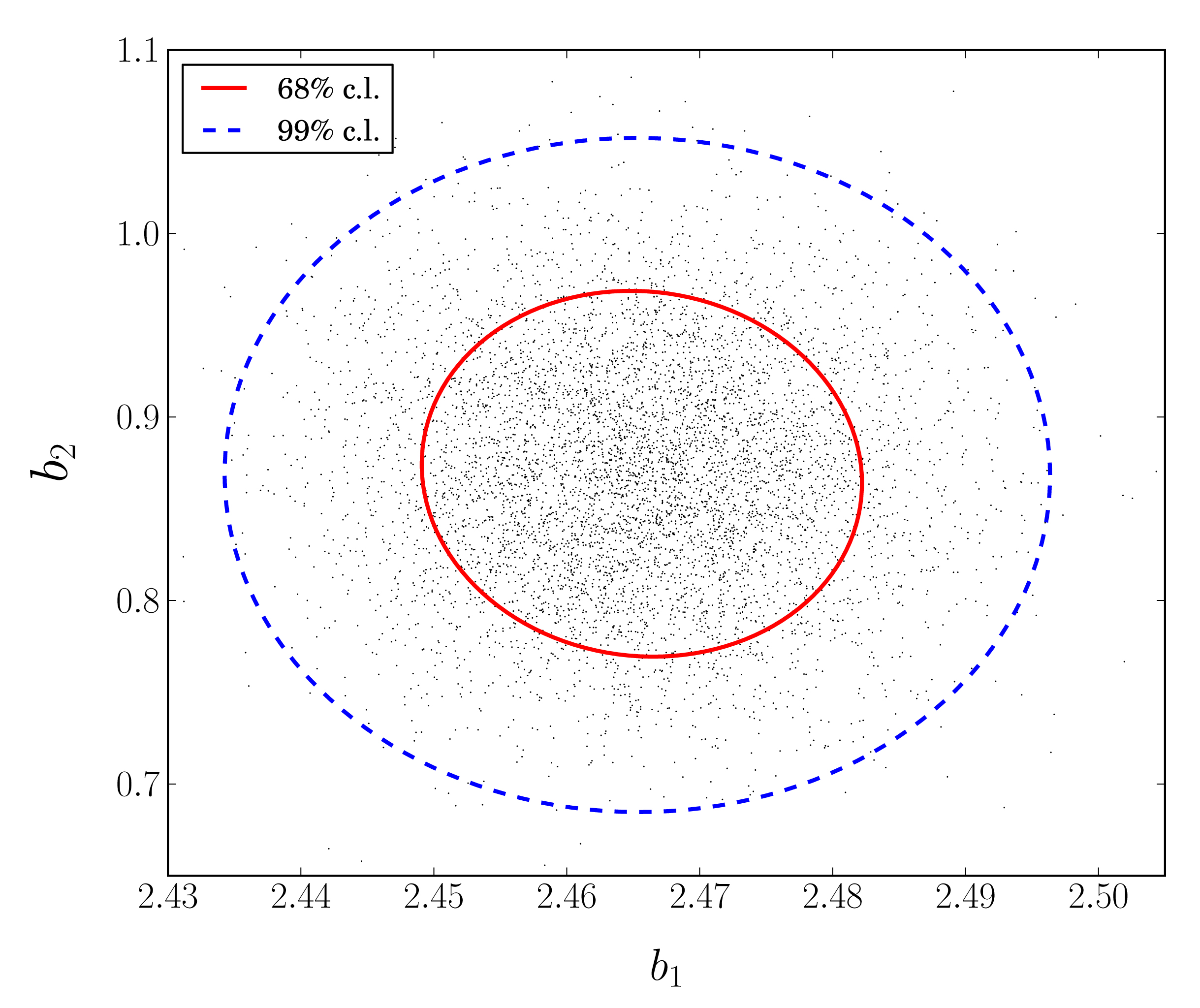} 
\includegraphics[scale=0.35]{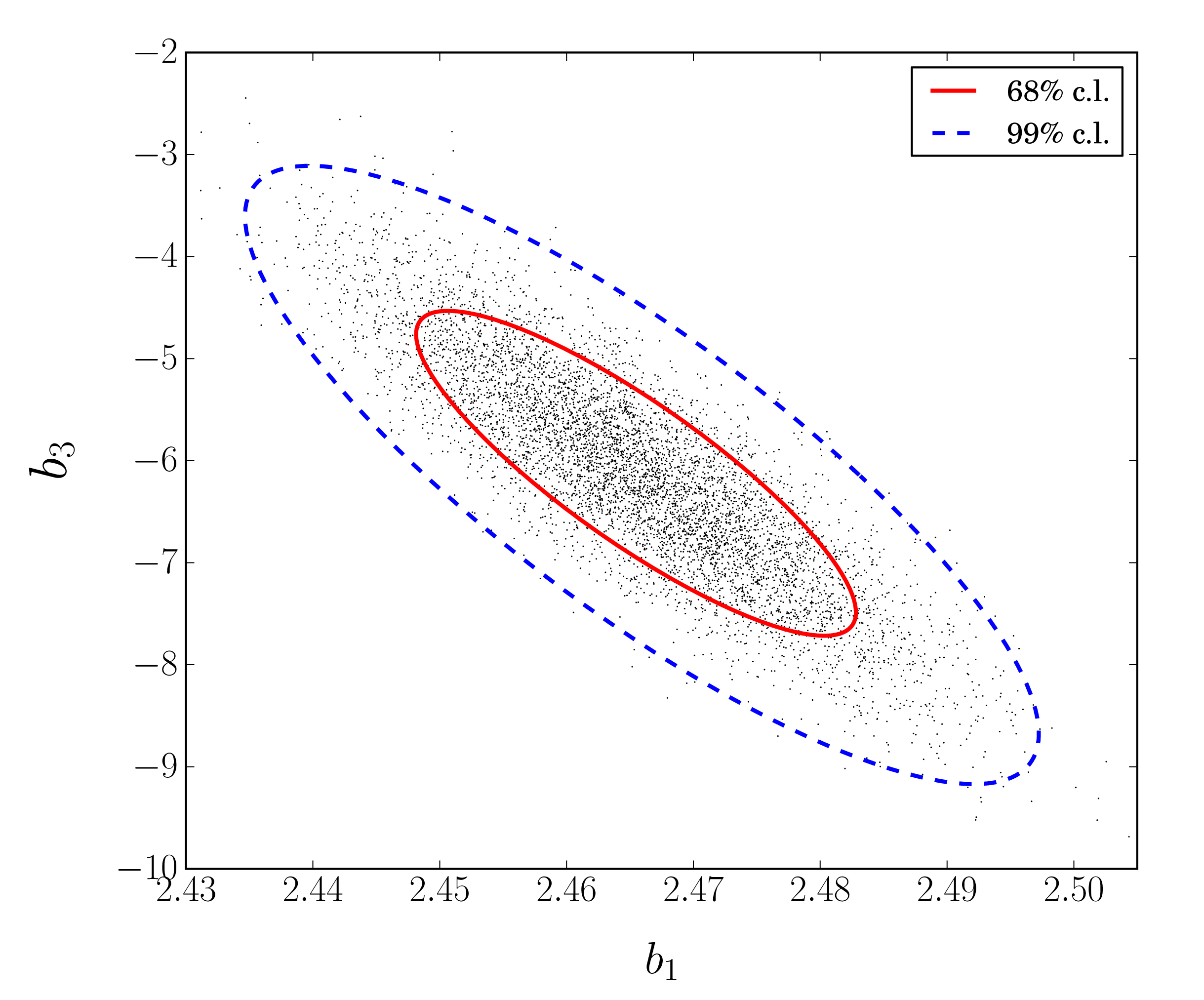}
\includegraphics[scale=0.35]{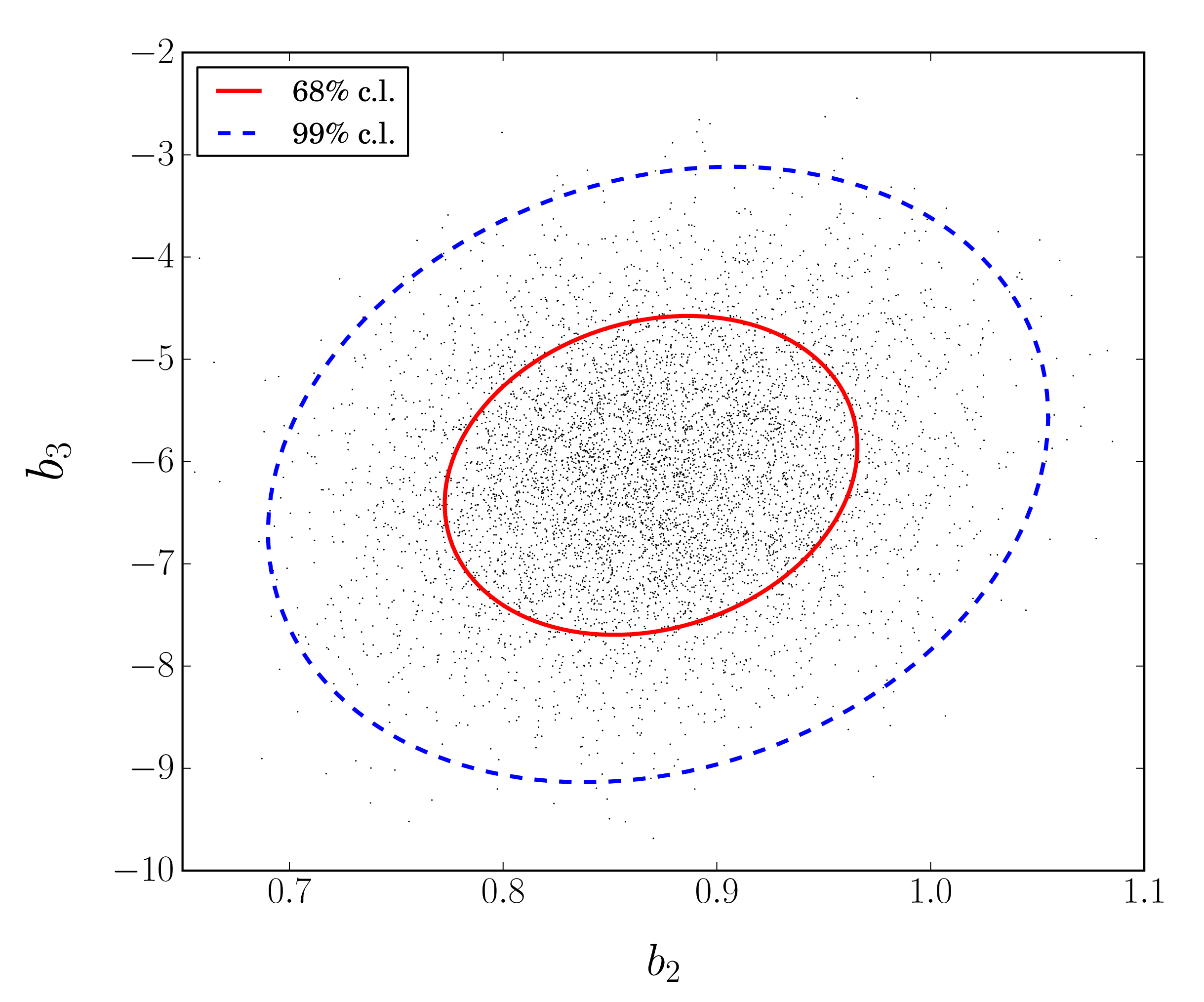}
\caption{Scatter plots of the results obtained with each bootstrap realization in the planes $b_1-b_2$, $b_1-b_3$ and $b_2-b_3$ at $\log M = 14.15$ and $z=0$.  The contours correspond to $68\%$ (red) and $99\%$ (dashed blue) confidence levels.}
\label{fig:contours}
\end{figure}

The bias parameters obtained from separate universe simulations for a given halo sample are in general significantly correlated.  From the covariance matrix of the measured Lagrangian bias parameters, we found that $b_n^L$ is correlated with $b_{n+2}^L$; this propagates to the Eulerian bias.  The correlation coefficients for the Eulerian bias parameters are only weakly dependent on the mass and read $\rho(b_1,b_2)\approx 0.0\pm0.15,\, \rho(b_1,b_3)\approx -0.80\pm0.05$ and $\rho(b_2,b_3)\approx 0.20\pm0.20$ at $z=0$. Note that the $b_2-b_3$ correlation is the most mass dependent with $\rho(b_2,b_3)\approx 0$ at low mass and $\approx 0.4$ at high mass.

To further illustrate this covariance between the measurements, we show in \reffig{contours} scatter plots of the results obtained with each bootstrap realization in the planes $b_1-b_2$, $b_1-b_3$ and $b_2-b_3$. We present the results at $\log M = 14.15$ and $z=0$ . Each time, we also trace out the contours corresponding to the $68\%$ and $99\%$ confidence levels. The covariance between $b_1$ and $b_3$ is clearly significant, illustrating the general trend that $b_n$ is highly correlated with $b_{n+2}$.

%%%%%%%%%%%%%%%%%%%%%%%%%%%%%%%%%%%%%%%%%%%%%%%%%%%%%%%%%%%%%%%%%%%%%%%%%%%%
%%%%%%%%%%%%%%%%%%%%%%%%%%%%%%%%%%%%%%%%%%%%%%%%%%%%%%%%%%%%%%%%%%%%%%%%%%%%
\bibliography{references}
\end{document}